\documentclass[sigconf]{acmart}
\usepackage{algorithmic}
\usepackage{graphicx}
\usepackage{textcomp}
\usepackage{graphicx}
\usepackage{caption}
\usepackage{subcaption}
\usepackage{bbm}
\usepackage{color}
\usepackage{hyperref}
\usepackage{multirow}
\usepackage{enumitem}
\usepackage{algorithm} 
\usepackage{algorithmic}
\usepackage{enumerate}
\usepackage{amsmath}
\usepackage{booktabs}
\usepackage{multirow}
\usepackage{stmaryrd}
\usepackage{mathrsfs} % needed for mathscr
\usepackage{color}
\usepackage{adjustbox}
\usepackage{balance}

\newcommand{\eg}{\textit{e.g.}}

\newcommand{\ie}{\emph{i.e.}}

\setlist[itemize]{leftmargin=*}
\renewcommand\footnotetextcopyrightpermission[1]{}
\AtBeginDocument{%
  }

\setcopyright{acmlicensed}
\copyrightyear{2018}
\acmYear{2018}
\acmDOI{XXXXXXX.XXXXXXX}

\acmConference[Conference acronym 'XX]{Make sure to enter the correct
  conference title from your rights confirmation emai}{June 03--05,
  2018}{Woodstock, NY}
\acmBooktitle{Woodstock '18: ACM Symposium on Neural Gaze Detection,
 June 03--05, 2018, Woodstock, NY} 
\acmISBN{978-1-4503-XXXX-X/18/06}

\begin{document}

\title{Bursting Filter Bubble: Enhancing Serendipity Recommendations with Aligned Large Language Models}

 
\author{Yunjia Xi$^{1*\dag}$, Muyan Weng$^{1*\dag}$, Wen Chen$^{2\ddag}$, Chao Yi$^2$, Dian Chen$^2$, Gaoyang Guo$^2$, Mao Zhang$^2$,\\Jian Wu$^2$, Yuning Jiang$^2$, Qingwen Liu$^2$, Yong Yu$^1$, Weinan Zhang$^{1\ddag}$}
% \affiliation{$^1$Shanghai Jiao Tong University, $^2$Alibaba Group}
\affiliation{
  \institution{$^1$Shanghai Jiao Tong University, $^2$Alibaba Group}\country{}
}
\email{{chenyu.cw,yichao.yc,joshua.cd,guogaoyang.ggy,yizhang.zm,joshuawu.wujian,mengzhu.jyn,xiangsheng.lqw}@alibaba-inc.com,{xiyunjia, 1355029251, yyu, wnzhang}@sjtu.edu.cn}
% \email{}

% \author{Yunjia Xi}
% % \email{xiyunjia@sjtu.edu.cn}
% \affiliation{%
%   \institution{Shanghai Jiao Tong University}
%   \city{Shanghai}
%   \country{China}
% }

% \author{Muyan Weng}
% \email{1355029251@sjtu.edu.cn}
% \affiliation{%
%   \institution{Shanghai Jiao Tong University}
%   \city{Shanghai}
%   \country{China}
% }

% \author{Weinan Zhang}
% \authornote{The corresponding author.}
% \email{wnzhang@sjtu.edu.cn}
% \affiliation{%
%   \institution{Shanghai Jiao Tong University}
%   \city{Shanghai}
%   \country{China}
% }

% \author{Yong Yu}
% \authornotemark[2]
% \email{yyu@sjtu.edu.cn}
% \affiliation{%
%   \institution{Shanghai Jiao Tong University}
%   \city{Shanghai}
%   \country{China}
% }

\renewcommand{\shortauthors}{Yunjia Xi et al.}

\begin{abstract}
  Recommender systems (RSs) often suffer from the feedback loop phenomenon, \ie, RSs are trained on data biased by their recommendations. This leads to the filter bubble effect that reinforces homogeneous content and reduces user satisfaction. To this end, serendipity recommendations, which offer unexpected yet relevant items, are proposed. Recently, large language models (LLMs) have shown potential in serendipity prediction due to their extensive world knowledge and reasoning capabilities. However, they still face challenges in aligning serendipity judgments with human assessments, handling long user behavior sequences, and meeting the latency requirements of industrial RSs. To address these issues, we propose SERAL (\underline{Se}rendipity \underline{R}ecommendations with \underline{A}ligned \underline{L}arge Language Models), a framework comprising three stages: (1) \textbf{Cognition Profile Generation} to compress user behavior into multi-level profiles; (2) \textbf{SerenGPT Alignment} to align serendipity judgments with human preferences using enriched training data; and (3) \textbf{Nearline Adaptation} to integrate SerenGPT into industrial RSs pipelines efficiently. Online experiments demonstrate that SERAL improves exposure ratio (PVR), clicks, and transactions of serendipitous items by 5.7\%, 29.56\%, and 27.6\%, enhancing user experience without much impact on overall revenue. Now, it has been fully deployed in "Guess What You Like" of Taobao App homepage.
\end{abstract}
% SERAL offers a novel solution for deploying LLM-based serendipity recommendations in large-scale RSs while addressing filter bubble challenges.

%    <concept>
%        <concept_id>10002951.10003317.10003347.10003350</concept_id>
%        <concept_desc>Information systems~Recommender systems</concept_desc>
%        <concept_significance>500</concept_significance>
%        </concept>
%  </ccs2012>

% \ccsdesc[500]{Information systems~Recommender systems}

% \keywords{Recommender System, Large Language Model, Knowledge Augmentation}


% \maketitle

\maketitle

\let\thefootnote\relax\footnotetext{* Equal contribution}
\footnotetext{\dag  Work done while interning at TaoBao}
\footnotetext{\ddag 
 Corresponding author}
\vspace{-5pt}
\section{Introduction}

% Introduce filter bubble

% - cause: the feedback loop

% - harm: user boredom, decrease user satisfaction 

% - possible solution: Only increasing diversity and novelty can mitigate homogenization, but it may also introduce irrelevant content, harming the user experience. A more effective approach is serendipity recommendation, aiming to recommend items that provide users with unexpected yet appealing experiences. A serendipitous item should exceed users' expectations while still capturing their interest, encompassing factors like unexpectedness, novelty, diversity, and relevance.
% \newline

\begin{figure}
    \centering
    \vspace{-5pt}
    \includegraphics[trim={0.3cm 0.5cm
    0 0},clip,width=0.5\textwidth]{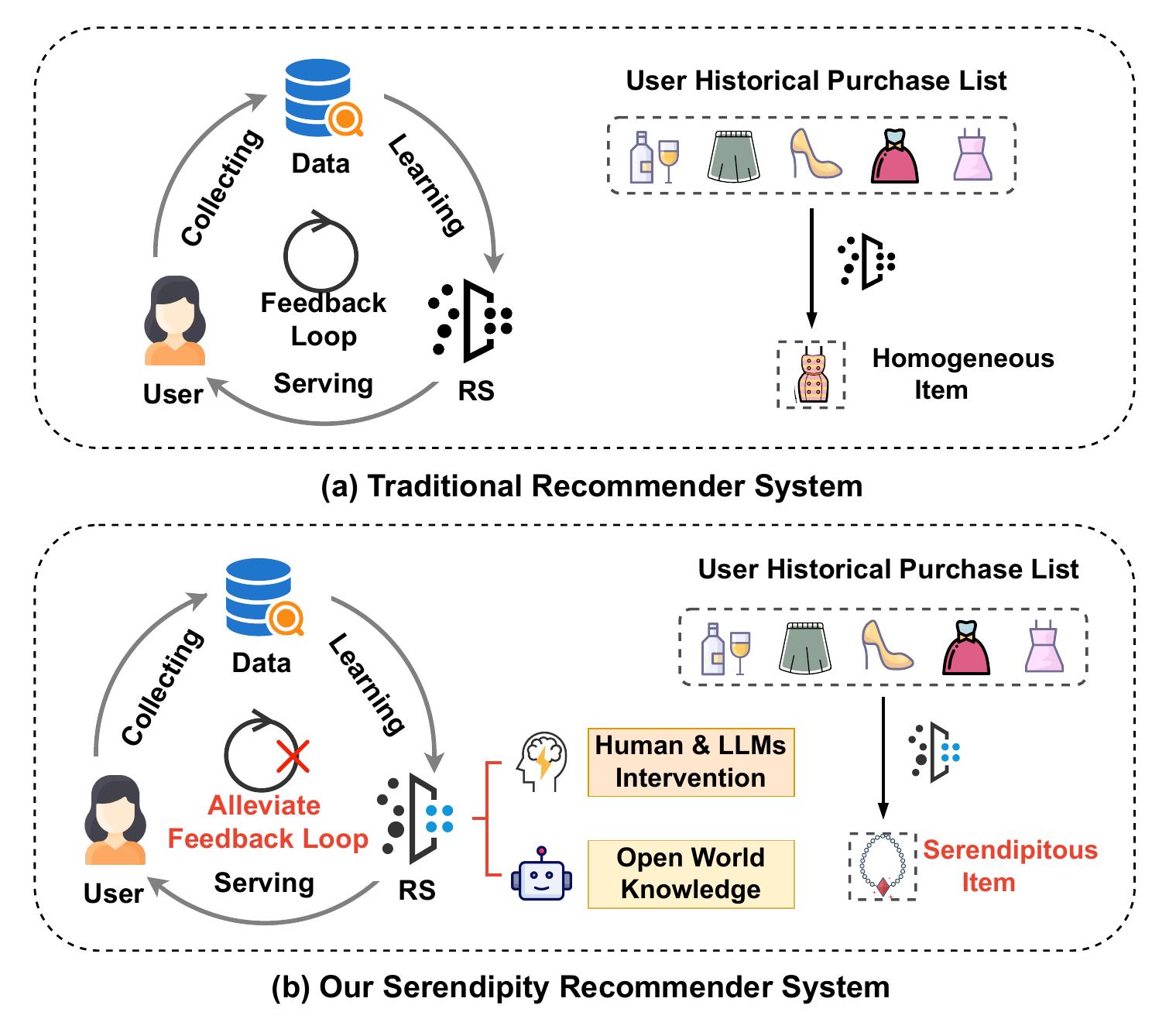}
    \vspace{-20pt}
    \caption{The comparison between (a) traditional recommender system and (b) our serendipity recommender system.}
    \vspace{-15pt}
    \label{fig:compare_intro}
\end{figure}

Nowadays, recommender systems (RSs) play an indispensable role in alleviating information overload and delivering personalized recommendations~\cite{fu2023deep,li2023breaking}. These systems predominantly rely on deep learning algorithms trained on users' historical interaction data and emphasize user preferences~\cite{zhou2018deep,xi2022multi,qin2020user}. However, due to the \textit{feedback loop} phenomenon, the content users encounter is largely determined by the system's prior recommendations, in turn narrowing down the training data that RSs can utilize. In other words, RSs are trained on data biased by its recommendations~\cite{li2023breaking,krauth2022breaking,mansoury2020feedback,chen2023bias}, as shown in Figure~\ref{fig:compare_intro} (a). The feedback loop, coupled with an over-specialization in user recent preferences, drives the system's recommendations and user behavior to converge around a narrow subset of homogeneous content, commonly called the \textit{filter bubble} effect~\cite{fu2023deep,aridor2020deconstructing,nguyen2014exploring}. As a result, RSs will persistently offer users familiar, homogeneous content, leading to user boredom and significantly diminishing user satisfaction~\cite{fu2023deep,niu2021luckyfind,niu2018adaptive}.

To address the challenges posed by the filter bubble, serendipity recommendation is advocated by many researchers, which aims to recommend items that provide users with unexpected yet appealing experiences~\cite{fu2023deep,ziarani2021serendipity,kotkov2016survey}. A serendipitous item should exceed users' expectations and counteract content homogenization while still capturing their interest,  which encompasses two key factors:  unexpectedness and relevance~\cite{fu2023deep,kotkov2024overview}. In recent years, serendipity recommendations have predominantly been driven by deep learning techniques~\cite{fu2023deep,fu2023wisdom,li2020directional,li2020purs,zhang2021snpr,wang2023industrial,cheng2017learning,liu2023personalized}, and have achieved promising results. However, due to the scarcity of serendipity data, which often relies on user surveys, they have to adopt smaller models and depend on biased recommendation data for augmentation. This may reinforce feedback loops, making it challenging to burst filter bubbles and accurately identify serendipitous items.

Recently, large language models (LLMs) 
have showcased remarkable proficiency in various domains, strikingly close to human-level performance~\cite{bubeck2023sparks,achiam2023gpt}. Notably, LLMs possess extensive world knowledge and exhibit remarkable comprehension and reasoning abilities. This enables them to incorporate external knowledge and intervention into RSs, break the feedback loop, and identify serendipitous items effectively, as shown in Figure~\ref{fig:compare_intro} (b). Moreover, LLM-based recommenders demonstrate sample efficiency and can attain superior performance with limited training data~\cite{lin2025large}. Preliminary work has attempted to apply LLMs to serendipity recommendations and achieve considerable results~\cite{tokutake2024can,fu2024art}, but they still suffer from the following issues, which limit their effectiveness and deployment.

% However, preliminary work~\cite{tokutake2024can,fu2024art} finds that while LLMs' performance is comparable to baselines, their serendipity assessments do not have a high agreement rate with humans. Therefore, we aim to improve LLMs' ability to recommend serendipitous items by addressing the issues below.

\textit{Firstly, serendipity judgments from LLMs are not aligned with human assessments.} Preliminary work leverages LLMs to evaluate item serendipity in a zero-shot manner and finds LLMs' judgments are not well-aligned with human assessments~\cite{tokutake2024can}. Serendipity relies heavily on personal context -- what feels serendipitous to one user may not to another~\cite{kotkov2024overview}. LLMs without fine-tuning do not develop a deep understanding of individual user preferences. Moreover, human perceptions of serendipity involve relevance and unexpectedness~\cite{fu2023deep}. Prompts may not be sufficient to guide LLMs toward accurately identifying the nuanced and subjective nature of serendipity because they offer limited control over how the model interprets unexpectedness and relevance, leading to misalignment with human expectations. Therefore, without dedicated alignment, LLMs may lack the tailored understanding of users and an accurate grasp of serendipity to align with human judgments.

\textit{Secondly, the length of user history that LLMs can utilize is limited, which poses a challenge in achieving a deep understanding of users.} SOTA recommendation models show that long user sequences are crucial for understanding user preferences~\cite{hou2024cross,si2024twin,feng2024context}. Similarly, serendipity recommendation relies on sufficient user history to identify serendipity, as it hinges on a nuanced understanding of an individual’s unique interests, surprises, and discoveries. However, LLMs struggle to comprehend long user behavior sequences, even when those sequences fall well within their context window~\cite{lin2024rella}. Studies exploring LLMs for serendipity also indicate their performance peaks with medium-length user sequences~\cite{tokutake2024can}. 
Given that user behavior histories in RSs are often considerably longer, even reaching 10,000~\cite{hou2024cross,si2024twin}, LLMs' inability to incorporate long sequences brings an insufficient understanding of the user’s nuanced preferences, obstructing them from identifying serendipitous items.

% \textit{Secondly, the length of user history that LLMs can utilize is limited, which poses a challenge in achieving a deep understanding of users.} Serendipity hinges on a nuanced grasp of an individual’s unique interests, surprises, and discoveries -- areas where LLMs fall short without sufficient personalized context or interactions. However, previous studies reveal that LLMs struggle to comprehend long user behavior sequences, even when those sequences fall well within their context window~\cite{lin2024rella}. Studies exploring LLMs for serendipity indicate their performance peaks with medium-length user sequences~\cite{tokutake2024can}.  Given that user behavior histories in RSs are often considerably longer, LLMs' inability to incorporate long sequences results in an insufficient understanding of the user’s nuanced preferences. Consequently, this limitation hinders their capacity to predict serendipitous items accurately.

\textit{Lastly, the current LLM-based serendipity recommendation poses challenges for online serving, making it difficult to address filter bubble issues in industrial RSs.} In existing approaches~\cite{tokutake2024can,fu2024art}, LLMs must assess each item's serendipity based on users' behavior sequences. However, industrial RSs handle vast numbers of users and items. Even if filtering down candidate items, its cost remains substantial. Furthermore, these methods necessitate high-latency online inference with LLMs, which fails to meet the latency requirements of industrial systems within hundreds of milliseconds~\cite{xi2023device}. Thus, deploying LLM-based serendipity recommendations efficiently and efficiently within industrial RSs remains a pivotal issue.

% - LLMs' serendipity judgments are not aligned with human assessments. 

% - The length of user history that LLMs can utilize is limited~\cite{tokutake2024can}. 

% - The current approach, which directly uses LLMs to assess an item's serendipity, poses challenges for online serving, making it difficult to solve filter bubble issues in industrial RSs.
% \newline

Therefore, we propose \underline{Se}rendipity \underline{R}ecommendations with \underline{A}ligned \underline{L}LMs (\textit{SERAL}). First, \textbf{Cognition Profile Generation} mimics human cognitive processes and compresses user information and long behavior sequences into concise, multi-level cognitive profiles, \eg, static, short-term, and long-term profiles. This prevents LLMs from handling long behavior sequences. Next, \textbf{SerenGPT Alignment} trains an LLM-based recommender, SerenGPT, to generate next serendipity items and aligns its serendipity judgments with human assessments through preference alignment algorithm IPO~\cite{azar2024general}. During the training data selection, we incorporate rich external information, including the open-world knowledge of LLMs and collaborative data intervention (CDI) between LLMs and humans, to break the feedback loop from biased RSs data. Lastly, in \textbf{Nearline Adaptation}, we design a nearline serendipity channel to produce and cache high-quality candidate items with SerenGPT. Then, cached candidates are incorporated into a traditional multi-stage recommendation pipeline (online personalized channel) at the ranking stage, avoiding the efficiency issues of LLMs in online serving. Our contributions can be summarized as follows:
\begin{itemize}
    \item We identify LLMs' advantages to detect serendipitous items and propose \textbf{SERAL}. To the best of our knowledge, this is the \textit{first work that utilizes and deploys LLMs for serendipity recommendation, addressing the filter bubble issue in industrial RSs}.
    % \item We devise three stages to enhance LLMs' ability to identify serendipity and feasibility to deployment, with \textbf{Congnition Profile Generation} to compress long user behaviors into multi-level cognitive profiles, \textbf{SerenGPT Alignment} to align the LLM-based serendipity recommender with human assessments via DPO, and \textbf{Nearline Adaptation} stage to incorporate SerenGPT into traditional multi-stage RSs efficiently.
    \item We devise three stages to enhance LLMs' ability to identify serendipity and feasibility to deployment, with Cognition Profile Generation to overcome LLMs' limitation in handling long sequences, SerenGPT Alignment to align LLMs with human serendipity judgments via IPO, and Nearline Adaptation stage to address deployment challenges.
    \item SERAL has been \textbf{fully deployed online} and improves serendipity exposure ratio (PVR), clicks, and transactions by 5.7\%, 29.56\%, and 27.6\%, without much impact on overall revenue. 
    \item Through \textbf{long-term online experiments}, we first validate that enhancing serendipity in large-scale RSs can notably improve user experience (3.04\% improvement in UV3) and even increase revenue (0.98\% increase in transaction volume).
\end{itemize}

\section{Related Work}
\subsection{Serendipity Recommendations}
% The concept of serendipity, how it solves filter bubbles, and its distinction with diversity and novelty (ref ~\cite{fu2024art})

Due to feedback loops, traditional RSs tend to recommend items similar to what users already know, leading to user boredom. To address this, serendipity is introduced in RSs, referring to unexpected yet appealing experiences~\cite{fu2023deep,ziarani2021serendipity,kotkov2016survey}. According to the Britannica Dictionary, serendipity is "luck that takes the form of finding valuable or pleasant things that are not looked for." Although there is no universally agreed-upon definition of recommendation serendipity, most research agrees that it should involve two central aspects: unexpectedness and relevance~\cite{fu2024art,fu2023deep,fu2023wisdom}, \ie, 
it should go beyond user expectations while also engaging their interest. 
% These two aspects help RSs break out of the filter bubble of homogeneous recommendations, ultimately offering unexpected yet appealing experiences for users. 
Besides, diversity and novelty are two concepts often associated with serendipity~\cite{fu2023deep,kotkov2016survey,kaminskas2016diversity}. They can introduce unfamiliar items to the user, offering a fresh experience (unexpectedness). However, these items may not attract users (non-relevance) and even cause dissatisfaction~\cite{kotkov2016survey}. Thus, serendipity is more challenging, necessitating unexpected recommendations still attract users' interests.

% DL-based~\cite{fu2023deep} and LLM-based~\cite{tokutake2024can,fu2024art} serendipity recommendation

Deep learning-based serendipity recommendation can be categorized into pre-processing, in-processing, and post-processing approaches~\cite{fu2023deep}. The pre-processing stage typically involves data pre-processing, \eg, engineering serendipity features~\cite{li2019haes,kim2019sequential}. 
% For instance, HAES~\cite{li2019haes} leverages the users’ demographic information and movies’ statistical information to calculate the elasticity to discover serendipitous movies. 
The in-processing approaches incorporate serendipity during model training and learn a serendipity representation~\cite{fu2023wisdom,li2020directional,li2020purs,zhang2021snpr,wang2023industrial}.
% For example, SerenEnhance~\cite{fu2023wisdom} learning the fine-grained facets of serendipity during training. 
Post-processing methods re-rank given relevance-oriented lists to generate serendipity-oriented recommendation lists~\cite{cheng2017learning,liu2023personalized}. 
% For instance, DCF~\cite{cheng2017learning} re-ranks candidate items to maximize diversity scores. 
In-processing is generally the most prevalent and effective. However, due to the scarcity of serendipity data, which often relies on user annotations, they must adopt smaller models and depend on biased data for augmentation. This may reinforce feedback loops, making bursting filter bubbles and identifying serendipity challenging.

Recently, LLMs have been applied to serendipity recommendations~\cite{fu2024art,tokutake2024can}. For example, in~\cite{tokutake2024can}, LLMs evaluate items' serendipity in a zero-shot manner. SerenPrompt~\cite{fu2024art} designs various prompts for LLMs to assess serendipity. These preliminary attempts show the promising potential of LLMs in serendipity recommendations. Still, several unresolved challenges remain, such as aligning LLMs with humans in serendipity and their deployment in industrial RSs.

\subsection{LLM-based Recommendations}
The past few years have witnessed a growing interest in LLM-based RSs~\cite{li2023large,zhu2023large,chen2023large,fan2023recommender,wu2023survey,liu2023pre,yu2023self,lin2023can}, which can be roughly classified into two categories: (1) LLMs as recommenders and (2) LLMs as components of traditional recommenders. The former is to adopt LLMs as recommenders to generate recommendations directly~\cite{xi2024memocrs,chatrec,xi2024play,lin2024rella,bao2023tallrec,zheng2024adapting,zhu2024collaborative,zheng2024harnessing,dong2024unsupervised,tan2024idgenrec,wu2024coral,kim2024large,wang2024eager,lin2024bridging}. Early attempts primarily utilize zero-shot LLMs~\cite{chatrec,xi2024memocrs}. 
% For instance, ChatRec~\cite{chatrec} and MemoCRS~\cite{xi2024memocrs} employ LLMs as recommender system interfaces for conversational multi-round recommendations. 
However, this falls behind SOTA recommendation models, so the focus of later work shifts to how to inject recommendation knowledge into LLMs, primarily through fine-tuning, such as TALLRec~\cite{bao2023tallrec} fine-tuned on LLaMA-7B~\cite{llama}. 
% ReLLa~\cite{lin2023rella} design retrieval-enhanced instruction tuning by adopting semantic user behavior retrieval as a data augmentation technique and finetunes Vicuna-13B.
The latter integrates knowledge from LLMs into conventional RSs to enhance the recommendation performance and meanwhile avoid the online inference latency issue of LLMs~\cite{xi2023towards,lyu2023llm,ren2024representation,xi2024efficient,xi2024decoding,liu2024once,ren2024enhancing,du2024disco,tian2024reland,wang2024can}. For example, KAR~\cite{xi2023towards} and RLMRec~\cite{ren2024representation} integrate open-world knowledge from LLMs into RSs. 
% Some researchers propose S\&R Multi-Domain Foundation model~\cite{gong2023unified}, which finetunes LLMs on recommendation and search data to extract domain invariant features for promoting performance in cold-start scenarios. 

% There are two categories: LLMs as recommenders and LLMs as feature enhancers. The latter is more suitable for deployment.

The latter is more deployable for industrial RSs when integrating RSs and LLMs~\cite{xi2023towards,zhang2024notellm,xi2024decoding,luo2024kellmrec}. The inference latency of LLMs is very high, and current industrial RSs cannot support the online inference of LLMs that the first approach requires. The latter only involves LLMs offline to generate features, which minimizes its impact on online inference latency. However, these deployable methods have only focused on improving recommendation accuracy and may struggle to adapt to data distribution changes. Therefore, we focus on serendipity recommendations and adopt nearline caching solutions to avoid LLMs' online inference while ensuring rapid updates to adapt to distribution changes.
% , efficiently integrating LLMs into industrial RSs. 
% To this end, our work follows the latter but shifts the focus to serendipitous recommendations, incorporating serendipitous items generated by LLMs with traditional RSs.
\section{Preliminaries}\label{sect:network_structure}
\subsection{Task Formulation}
Given an item set $\mathcal{I}$ and user set $\mathcal{U}$, for each user $u\in\mathcal{U}$ with history $H_u=\{h_1,\ldots,h_m\}$ of length $m$, where $h_i, i=1,\ldots,m$ denotes the $i$-th item the user interacted recently, the goal of serendipity recommendation is to find an item which is serendipity to the user $u$. Previous approaches convert it into a matching problem~\cite{fu2023wisdom,fu2024art}, where each candidate item $i$ is scored based on $H_u$, \ie,
\begin{equation}
    y_{u,i} = matching(i, H_u),
\end{equation}
where $y_{u,i}$ denotes the predicted serendipity score and $matching(\cdot)$ is matching function.
However, it is highly inefficient for LLMs to score all items. Therefore, we can formulate it as a generation problem, generating the next serendipitous item $i$ based on $H_u$:
\begin{equation}
    i = \pi_{\theta}(H_u),
\end{equation}
where $i$ is the predicted serendipity item generated by LLM $\pi_{\theta}(\cdot)$.

\subsection{Serendipity in industrial RSs}
Identifying serendipity is usually a challenging issue. Previous serendipity datasets often relied on user surveys~\cite{kotkov2018investigating,chen2019serendipity}. In industrial scenarios, vast amounts of interaction data are generated daily, making user surveys impractical. Therefore, by combining the key aspects of serendipity, unexpectedness, and relevance, we propose an industrial and simple definition of serendipity: 
\newtheorem{defi}{Definition}
\begin{defi}
A serendipitous item is identified by two criteria: it must be clicked, and its atomic category should not have been present in the user's valid exposures over the last N days with visits.
\label{def:serend} 
\end{defi}

In e-commerce, item classifications typically follow a tree structure, and we refer to the indivisible subcategories at the leaf nodes as atomic categories. The unexposed category guarantees the user's sense of freshness and surprise, while the click indicates its relevance. Here, $N$ may vary with scenarios; in our scenario, we set $N=10$ and find that optimizing this metric significantly improves user experience. However, such serendipity data is still scarce, and maintaining accuracy is still the primary goal of RSs. Therefore, the main goal of serendipity in industrial RSs is to maintain the system's utility (such as click-through rate (CTR), revenue) while increasing the proportion of serendipity items (\ie, S-PVR in Section~\ref{sec:metrics}), thereby keeping users engaged and intrigued. 
% Therefore, we need to consider adapting the serendipity model to the accuracy-oriented RSs, and we will discuss this in Section~\ref{sec:downstream}. 

\begin{figure*}
    \centering
    \vspace{-10pt}
    \includegraphics[trim={0.3cm 0.3cm 0 0},clip,width=\textwidth]{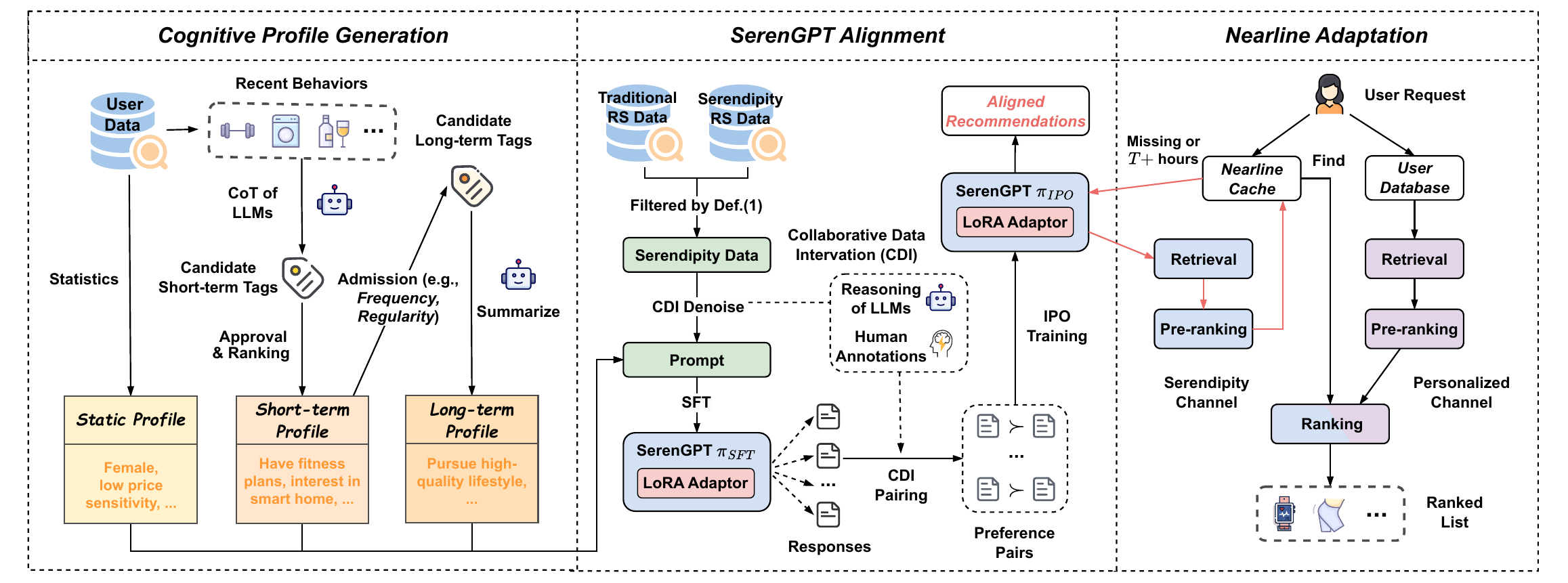}
    \vspace{-15pt}
    \caption{The overall framework of SERAL, with Cognitive Profile Generation to compresses long behavior sequences into concise profiles, SerenGPT Alignment to align LLMs' serendipity judgments with humans, and Nearline Adaptation to integrate serenGPT into industrial recommender systems efficiently.}
    \vspace{-10pt}
    \label{fig:fw}
\end{figure*}

\section{Methodology}
\subsection{Overview}
As illustrated in Figure~\ref{fig:fw}, the workflow of our proposed SERAL encompasses three stages. Firstly, \textbf{Cognitive Profile Creation} mirrors human cognitive processes and compresses the user's long behavior sequences into concise cognitive profiles, including static, short-term, and long-term profiles. Secondly, in \textbf{SerenGPT Alignment}, we devise an LLM-based serendipity recommender, SerenGPT, to predict next serendipitous items. Then, we train and align SerenGPT's judgments with humans via a preference alignment method IPO~\cite{azar2024general}. When curating training data, we incorporate Collaborative Data Intervention (CDI) from LLMs and humans, breaking the internal feedback loop of RSs. Finally, in \textbf{Nearline Adaptation}, we implement a serendipity channel with SerenGPT and integrate its recommendations into accuracy-oriented RSs via nearline cache. This solves LLMs' efficiency issues of online serving.

\subsection{Cognition Profile Generation}
User historical behavior is critical for modeling serendipity; however, LLMs struggle to leverage long user behavior sequences. Previous research has shown that long historical sequences, whose lengths often reach up to 10,000~\cite{pi2020search,hou2024cross,si2024twin,feng2024context} in large-scale RSs, contain a wealth of user preferences and could enhance recommendation~\cite{pi2020search,qin2020user}.  Serendipity relies heavily on capturing unique user preferences, surprises, and discoveries, which also requires a deep understanding of longer behavior sequences. Nevertheless, LLMs struggle to comprehend long user behaviors, with performance declining well before reaching the context window of LLMs~\cite{lin2024rella}, obstructing them from identifying serendipitous items. 
% Therefore, it is crucial to improve LLMs' ability to grasp user interests and intentions from long behavior sequences for serendipity recommendations.

To address this, we design cognition profile generation, which compresses long user behavior into precise profiles that LLMs can easily recognize. This process mimics human cognition by first understanding foundational information to form a static profile, then capturing short-term intent from recent behaviors, and finally summarizing short-term intent from long-term patterns, creating multi-layer cognitive profiles that LLMs can effectively utilize.

\subsubsection{Static Profile} This profile contains fundamental information, encompassing basic information and statistical characteristics. These features are generally stable and do not require further processing, as they can be directly retrieved from databases. For a given user $u$, we can convert these attributes into a textual profile $p_s^u$ with pre-defined templates.

% \subsubsection{Static Profile} This profile contains fundamental information, encompassing basic user information (\eg, age, occupation, gender, and location) and statistical characteristics (\eg, purchasing power and price sensitivity). These features are generally stable and do not require further processing or summarization, as they can be directly retrieved from databases. For a given user $u$, we can convert these attributes into a textual profile $p_s^u$ with pre-defined templates, such as \textit{``This user was born in \{\{year\}\}, is a \{\{gender\}\} working as a \{\{profession\}\}, resides in \{\{city\}\}, has a \{\{high/medium/low\}\} price sensitivity and a \{\{high/medium/low\}\} purchasing power.''}

\subsubsection{Short-term Profile} Short-term profile represents volatile, short-term intentions and interests extracted from the user's recent behaviors (\eg, search queries and titles of purchased products from the past 2-4 weeks). Many users' habits and interests are embedded within their behaviors. LLMs, with their strong reasoning abilities and extensive world knowledge~\cite{wei2022chain,llama}, can simulate human-like reasoning and extract users'  traits and interests from their behaviors. Thus, we first focus on refining profiles from recent behavior sequences via LLMs. Specifically, given recent behaviors $H_u$ of user $u$, LLMs infer candidate profile $p^u_{st}$ via Chain-of-Thought (CoT) reasoning~\cite{wei2022chain} based on prompt template $T_{st}$:
\begin{equation}
    p^u_{st} = f_{LLM}(H_u, T_{st}),
\label{eq:llm_recent}
\end{equation}
where $T_{st}$ possesses task description such as \textit{``Based on the user's purchasing history, infer their purchasing intent and generates diverse profile tags for the user.''} and format requirements like \textit{``The output format should be in JSON, with the user profile tag as key and the reasoning process as the value.''} To avoid irrelevant outputs and reduce hallucinations, we limit the user profile tags to preferred categories, brands, IPs, and attributes, possible interests, and possible usage scenarios. The full prompt can be found in Appendix~\ref{app:prompt}.

After the CoT reasoning of LLMs, we obtain several candidate profiles $p^u_{st}$. However, these profiles may be unreliable due to hallucinations. Therefore, an approval mechanism is necessary to calibrate a more accurate short-term profile. We maintain a tag pool for users to store credible tags. When new tags $p^u_{st}$ are generated, a historical prior check (whether it already exists in the tag pool) and a behavioral posterior check (based on the relevance between products' tags and profile tags) are employed to determine their credibility. Credible tags are then scored and ranked, with the higher-ranking tags selected as the short-term profile $\hat{p}^u_{st}$. 

% For example, we can determine whether a profile tag qualifies for the tag pool based on confidence or frequency and further score those profile tags via user behavior and tag priors, with the highest-ranked ones forming the user's recent profile $\hat{p}^u_{st}$.

% Next, LLMs infer the user's candidate intent based on the above keywords with chain-of-thought (COT) reasoning. 

% The admission criteria: The candidate intent is then scored and ranked after applying historical label priors, behavioral posteriors, and admission rules (confidence scoring).
\subsubsection{Long-term Profile}\label{sec:long-term} This refers to stable and long-term intentions derived from multiple short-term profiles. Compared to the short-term profile, the long-term profile focuses more on stability, accuracy, and long-term relevance, requiring a more rigorous reasoning and calibration process. To achieve this, we periodically mine the user's short-term profiles, with criteria like frequency, regularity, and long-term effectiveness as admission conditions, to get candidate long-term profiles $p^u_{lt}$. By combining these factors with the summarization and reasoning abilities of LLMs, we could refine the user's long-term profile $\hat{p}^u_{lt}$ as follows:
\begin{equation}
    \hat{p}^u_{lt} = f_{LLM}([p^u_{lt}, T_{lt}]),
\label{eq:llm_long}
\end{equation}
where $p^u_{lt}$ denotes candidate profiles that pass the admission criteria and $T_{lt}$ is a prompt template with task description like \textit{``Based on the given short-term profiles, summarize and refine the user's long-term profile.''} To ensure accuracy, long-term profiles also follow approval and ranking processes similar to short-term profiles.

Note that the LLMs employed for user profile generation are distinct from SerenGPT, the model discussed in section \ref{sec:align} for predicting serendipitous items. LLMs for cognitive profile generation are very powerful or highly specialized models for this task, \eg, advanced large-scale models at the
GPT-4 level or small LLMs fine-tuned on user profile data.

% LLMs are responsible for extracting and summarizing short-term profiles into candidate long-term intents.

% The admission criteria: frequency, periodicity, long-term effectiveness, and multi-dimensional verification. After confidence scoring, a decision is made on admitting the long-term intent.

\subsection{SerenGPT Alignment}\label{sec:align}
We now obtain cognition profiles, so LLMs can leverage profiles and other information to make serendipity recommendations. However, human perception of serendipity is nuanced and personalized, and zero-shot or even fine-tuned LLMs struggle to align with human perspectives on this~\cite{fu2024art,tokutake2024can}. To address this, we develop an LLM-based recommender, SerenGPT, trained with IPO~\cite{azar2024general}. In selecting data, we devise \textbf{Collaborative Data Intervention (CDI)}, which incorporates both intervention and annotations from LLMs and humans, along with the rich external knowledge of LLMs, to mitigate feedback loops and reduce the cost of human annotations.

\subsubsection{Serendipity Data Generation}\label{sec:data_gen}
% Diverse data sources: 

% - recommendation generated from relevance-based collaborative models. 

% - serendipity items directly recommended by LLMs (based on user behaviors, profiles, and external trends)

% the above data are filtered by specific rules 

% Combining GPT-4 with human annotation, we can identify serendipity items that can be inferred by profiles, history, and the current environment while excluding data that cannot be reasonably predicted, such as accidental clicks.

% A judgment model is trained on the above labeled data to select appropriate data for the upcoming SerenGPT training.

The filter bubble effect arises from RSs training on internal data biased by itself, leading to reinforced feedback loops. To break this cycle, the training of RSs should not solely rely on unmodified internal data; instead, it should integrate more external information and intervention. Therefore, we intervene in data collection by combining the world knowledge of LLMs and human annotations.
Our data sources primarily include:
\begin{itemize}
    \item Data from traditional accuracy-oriented recommender systems that align with the definition~\ref{def:serend} of serendipitous items.
    \item Data generated by serendipity recommender (\eg, SerenGPT or LLMs based on user profiles, historical behavior, and external trends) which also aligns with definition~\ref{def:serend}.
\end{itemize}

These two data sources may also contain some noise, such as accidental clicks. Thus, we design \textbf{Collaborative Data Intervention (CDI) Denoising}, employing powerful LLMs at GPT-4 level and human annotators to reduce noise. Specifically, the LLMs perform CoT reasoning~\cite{wei2022chain} based on user historical behavior and profiles to determine whether a target item is serendipitous for the user. Human annotators further assess the accuracy of the reasoning and conclusions from LLMs, generating more accurate serendipity labels. The annotated data, including user historical behaviors, profiles, target items, and serendipity labels, will be used to fine-tune a smaller-scale LLM (\eg, 7 billion parameters) to identify serendipity data for subsequent SerenGPT training. This allows for identifying serendipitous items with relatively low human annotation costs.
% The specific implementation of CDI here involves leveraging LLMs for annotations, followed by human evaluation to assess the rationality of annotations. 

\subsubsection{Supervised Fine-tuning (SFT)}\label{sec:sft}

Upon acquiring serendipity data, where each entry encompasses user $u$, historical behaviors $H_u$, and the next serendipitous item $i$, we proceed to fine-tune SerenGPT $\pi_\theta$. SerenGPT can be any causal language model that can analyze users' behaviors and profiles to generate the subsequent potential serendipitous item. First, the serendipity data are converted into a textual SFT dataset $D$, where each entry is a tuple $(x,y)$. Here, $y$ refers to the output, title of serendipitous item $i$, and $x$ represents the input prompt, \eg, ``\textit{Given the user's static profile $p^u_{s}$, short-term profile $\hat{p}^u_{st}$, long-term profile $\hat{p}^u_{lt}$, and recent historical behaviors $H_u$, generate the next serendipitous item that the user may be surprised by and enjoy.}'' To address delays in short-term profiles (which update every 15 days online), we also incorporate users' recent historical behaviors. Next, following Eq.~\eqref{eq:sft}, SerenGPT is supervised fine-tuned to augment its probability of generating desired responses.
\begin{equation}
    \begin{split}
    \mathcal{L}_{SFT}&=-E_{(x, y)\sim D}(\log \pi_\theta(y|x)) \\
    &=-E_{(x, y)\sim D}(\sum^{T}_{t=1}\log \pi_\theta(y_t|x,y_{1:t-1})).
\label{eq:sft}
\end{split}
\end{equation}

\subsubsection{Pair Generation and IPO Training}
In standard preference alignment frameworks, the LLM typically undergoes SFT, followed by preference alignment training on datasets comprising human-annotated preferences over pairs of model-generated responses. Techniques such as DPO~\cite{rafailov2024direct} are employed to align the model's outputs more closely with human preference. However, such methodologies encounter significant obstacles when applied to serendipity recommendations. Firstly, acquiring annotations directly from real users is impractical, while relying on other annotators may require substantial effort and could lead to inaccuracy. Secondly, our empirical findings indicate that LLMs after SFT tend to generate homogenized responses for serendipity recommendations, as shown in Appendix~\ref{app:diversity}. Even with ground truth recommendations as the preferred response, constructing a diverse preference dataset with discriminability between responses proves challenging.

To mitigate these limitations, we propose a \textbf{Collaborative Data Intervention (CDI) Pairing} mechanism, which synthesizes human expertise with teacher LLMs to curate high-quality preference pairs. Specifically, SerenGPT after SFT generates multiple recommendations by sampling, which are subsequently evaluated by teacher LLMs and assigned serendipity scores ranging from 1 to 6. Preference pairs are constructed by designating higher-scoring recommendations as preferred responses and lower-scoring ones as dis-preferred ones. For cases where scores are close, human annotations are performed to ensure accuracy. Here, the teacher LLMs employed in CDI are state-of-the-art, high-capability models at the GPT-4 level. These models enhance the quality and reliability of preference data, ensuring that the alignment training is rigorous and effective in capturing nuanced human preferences.

After obtaining the preference data, we train SerenGPT with the preference alignment algorithm IPO~\cite{azar2024general} as shown in Eq.~\eqref{eq:ipo}. The reason for choosing IPO over DPO~\cite{rafailov2024direct} is that during the training process, we observe that DPO tends to overfit, while the IPO algorithm helps alleviate this overfitting.

\begin{equation}
    \mathcal{L}_{IPO}=-E_{(x,y_w,y_l)\sim D'}\left[\left(\log\frac{\pi_\theta(y_w|x)\pi_{ref}(y_l|x)}{\pi_{ref}(y_w|x)\pi_{\theta}(y_l|x)} - \frac{\tau^{-1}}{2}\right)^2\right],
    \label{eq:ipo}
\end{equation}
where $D'$ denote the preference dataset,  $x$ denotes the input prompt, and $y_w$ and $y_l$ represent the preferred and dis-preferred responses. Here,  $\pi_{\theta}$ is the trainable LLM for preferences alignment, while $\pi_{ref}$ is a fixed reference policy, typically the original LLM. $\tau$ is the hyper-parameter of the regularization term, which balances learning human preferences and staying close to the reference policy, thereby preventing overfitting. In our experiments, we find that incorporating the SFT loss improves performance. Therefore, the final loss at this stage is:
\begin{equation}
    \mathcal{L} = \mathcal{L}_{IPO} +\alpha \mathcal{L}_{SFT}.
    \label{eq:final_ipo}
\end{equation}

\subsection{Nearline Adaptation}\label{sec:downstream} Nearline computation lies between online serving and offline processing. This concept was first introduced by Netflix~\cite{amatriain2013big,amatriain2013system} and has been used in many recommender systems~\cite{ma2024simple,borisyuk2024lignn,li2021truncation}. Nearline computation does model calculations at periodic intervals, such as daily or hourly, and caches the results for online serving. When a user request is initiated, the system combines the user's real-time information with the nearline cached results to generate recommendations. This approach improves recommender systems' efficiency and response speed while maintaining high accuracy.

The high latency and low real-time dependency of LLM-based serendipity recommendations make it suitable for the nearline approach. Firstly, LLMs have significant inference latency, which makes it hard to meet the low-latency demands of online serving in RSs, usually in 100 ms~\cite{xi2023device}. Secondly, serendipity recommendations do not rely much on real-time user behaviors. Its objective is to provide users with unexpected but valuable content, prioritizing long-term user satisfaction over real-time needs. Over-reliance on real-time behavior could reduce serendipity. Therefore, we adapt SerenGPT to large-scale RSs via a nearline approach. Specifically, we design two channels: an online personalized channel and a nearline serendipity channel, as shown in Figure~\ref{fig:fw}. The serendipity channel consists of SerenGPT, downstream models, and a nearline cache, while personalized channel is personalized multi-stage RSs.

% why adapt it to conventional models and why adapt it on both recall and ranking stage 
\subsubsection{Serendipity Channel}
This channel is responsible for leveraging SerenGPT, trained in the previous stage, for generating serendipity candidates and then caching the results in a nearline cache for online serving. Specifically, recommendations generated by SerenGPT first undergo multi-channel retrieval, such as keyword-based and vector-based retrieval. A pre-ranking model then processes the aggregated results, filtering out fewer but higher-quality candidates. The optimization of retrieval and pre-ranking models also considers serendipity. For instance, their training data includes a higher proportion of serendipity samples and incorporates optimization for hard negative samples related to serendipitous items. Finally, the serendipity candidates from the pre-ranking model are stored in a nearline cache. This cache retains those candidates within a time window of T hours, typically set to 24 hours, after which the cached candidates expire and are recalculated.

\subsubsection{Integration with Personalized Channel}

The personalized channel refers to traditional multi-stage recommender systems, without consideration for serendipity. In our dual-channel system, when a user triggers a recommendation, two pathways are activated: 
\begin{itemize}
    \item It collects the user's real-time behaviors and contextual information to generate personalized candidate items through the retrieval and pre-ranking stages of the personalized channel.
    \item It checks the nearline cache for the user's serendipity candidate items. The cached candidates are used directly if they are valid and unexpired (within T hours). If not, the system triggers the whole serendipity channel to generate new candidates, which are then added to the nearline cache.
\end{itemize}
Subsequently, serendipity candidates and personalized candidates are combined in the ranking stage. Since ranking is a part of the personalized channel and does not include optimizations for serendipity, we adjust the weights of serendipity candidates to make them more prominent, ensuring they get sufficient exposure. After the fusion in ranking, the top results are recommended to the user.

\section{Experiment}
\subsection{Setup}

\begin{table*}[h]

    \vspace{-10pt}
    \caption{Offline performance on serendipity recommendation.  The best result is given in bold, while the second-best value is underlined. The symbol * indicates statistically significant improvement over the best baselines( t-test with $p < 0.05$).}
    \vspace{-8pt}
    \centering
    \scalebox{0.95}{
    \setlength{\tabcolsep}{1.5mm}
    {\begin{tabular}{ccccc|cccc|cccc}
\toprule
\multirow{2}{*}{Method} & \multicolumn{4}{c|}{MAP$_{seren}$} & \multicolumn{4}{c|}{NDCG$_{seren}$} & \multicolumn{4}{c}{HR$_{seren}$} \\
\cmidrule{2-13} 
 & @1 & @3 & @5 & @10 & @1 & @3 & @5 & @10 & @1 & @3 & @5 & @10 \\
 \midrule
SASRec & 0.0606 & 0.0810 & 0.0863 & 0.0903 & 0.0606 & 0.0876 & 0.0971 & 0.1067 & 0.0606 & 0.1067 & 0.1294 & 0.1590 \\
Bert4Rec & 0.0273 & 0.0332 & 0.0339 & 0.0348 & 0.0273 & 0.0352 & 0.0364 & 0.0386 & 0.0273 & 0.0409 & 0.0439 & 0.0507 \\
\midrule
PURS & 0.0194 & 0.0247 & 0.0273 & 0.0297 & 0.0194 & 0.0267 & 0.0316 & 0.0377 & 0.0194 & 0.0323 & 0.0452 & 0.0645 \\
SerenEnhance & \underline{0.1900} & \underline{0.1956} & 0.1965 & 0.1972 & \underline{0.1900} & \underline{0.1974} & 0.1993 & 0.2007 & \underline{0.1900} & 0.2029 & 0.2074 & 0.2120 \\
\midrule
LLM4Seren & 0.0683 & 0.1163 & 0.1335 & 0.1544 & 0.0683 & 0.1192 & 0.1485 & 0.2047 & 0.0683 & 0.1893 & 0.2820 & 0.4965 \\
SerenPrompt & 0.0840 & 0.1750 & \underline{0.2187} & \underline{0.2455} & 0.0840 & 0.1904 & \underline{0.2725} & \underline{0.3531} & 0.0840 & \underline{0.3331} & \textbf{0.5723} & \textbf{0.8706} \\
\midrule
\textbf{SerenGPT} & \textbf{0.2861*} & \textbf{0.3260*} & \textbf{0.3395*} & \textbf{0.3509*} & \textbf{0.2861*} & \textbf{0.3353*} & \textbf{0.3587*} & \textbf{0.3847*} & \textbf{0.2861*} & \textbf{0.4625*} & \underline{0.5556} & \underline{0.6760}\\
 \bottomrule
\end{tabular}}
}
\label{tab:overall}
\vspace{-10pt}
\end{table*}

% \subsubsection{Scenario} Our experiments are conducted on Taobao, a leading large-scale e-commerce platform, including serendipitous product recommendations in "Guess What You Like" of Taobao homepage, and serendipitous search query predictions in Main Search.
% \begin{itemize}
%     \item \textbf{Product Recommendation} aims to predict next serendipitous item based on a user's historical behaviors and profile. For offline experiments, we filter out 19, 141 serendipity samples, splitting them into training and testing sets at 9:1. Since most baselines require scoring candidate item, we follow~\cite{fu2024art} and randomly select 99 negative items for each positive item in the test set to form a candidate set.  As our model SerenGPT generates the next item's title, we employ a matching algorithm to calculate the similarity between the generated title and the candidate items' titles, producing a ranking based on the match scores. For online serving, w
%     \item \textbf{Search Query Prediction} aims to predict the next serendipitous search query based on the user's historical behavior and profile.  The definition of serendipity remains consistent with Definition 1. For offline experiments, we filter out 26, 672 serendipity samples.
% \end{itemize}

\subsubsection{Scenario} Our experiments are conducted on \textbf{Taobao}, a leading large-scale e-commerce platform with billions of users and items. Our scenario is the "\textbf{Guess What You Like}" column of the Taobao App homepage and we aim to predict the next serendipitous item based on a user's historical behaviors and profiles. For offline experiments, we filter out 19, 141 serendipity samples from this scenario according to Section~\ref{sec:data_gen}, splitting them into training and testing sets at 9:1. Since most serendipity baselines~\cite{fu2024art,tokutake2024can} require ranking a small set of candidate items, we follow~\cite{fu2024art} and randomly select 99 negative (non-serendipitous) items for each positive (serendipitous) item in test set to form a candidate item set. As our model SerenGPT is a generative model that generates the item's title, we employ a matching algorithm to calculate the similarity between the generated title and the candidate items' titles, producing a ranking based on the match scores. 

\subsubsection{Baselines}
For offline experiments, we follow~\cite{fu2024art} and adopt three categories of baselines: \textbf{sequential recommendation, deep learning-based and LLM-based serendipity recommendation models}. Due to the scarcity of serendipity data, we incorporate non-serendipity click data for several baselines. Sequential recommendation algorithms, including \textbf{SASRec}~\cite{kang2018self} and \textbf{BERT4Rec}~\cite{sun2019bert4rec}, are trained on both serendipity and non-serendipity data. Deep learning-based serendipity recommendation methods are mainly trained on serendipity data, but also rely on non-serendipity data for augmentation. In this category, we chose two SOTA baselines \textbf{SerenEnhance}~\cite{fu2023wisdom} and  \textbf{PURS}~\cite{li2020purs}. LLM-based serendipity recommenders are trained only on serendipity data. This is a relatively new area, and we identify two baselines, \ie, \textbf{LLM4Seren}~\cite{tokutake2024can} and \textbf{SerenPrompt}~\cite{fu2024art}. Our model, SerenGPT, is also trained solely on serendipity data. More details can be found in Appendix~\ref{app:baselines}.

% \subsubsection{Baselines}
% For \textit{item prediction}, we consider sequential recommendation models \eg, \textbf{SASRec}~\cite{kang2018self} and \textbf{BERT4Rec}~\cite{sun2019bert4rec}, deep learning-based serendipity models like \textbf{SerenEnhance}~\cite{fu2023wisdom} and  \textbf{PURS}~\cite{li2020purs}, and LLM-based serendipity models, \eg, \textbf{LLM4Seren}~\cite{tokutake2024can} and \textbf{SerenPrompt}~\cite{fu2024art}. Due to the scarcity of serendipity data, many baselines incorporate regular click data. Similarly, we enhance the sequential recommendation and deep learning-based serendipity baselines with additional click data samples. For the \textit{search query prediction} task, since prior serendipity models do not address this task, we design several LLM-based baselines, including zero-shot LLMs (\textbf{ZSLLM}) and LLMs supervised fine-tuned on search query data (\textbf{SFT}), as well as different variants of preference alignment, \eg, \textbf{DPO}~\cite{rafailov2024direct}, \textbf{KTO}~\cite{ethayarajh2024kto}, and \textbf{SimPO}~\cite{meng2024simpo}, with then same backbone LLM and prompt as SerenGPT. More details can be found in Appendix~\ref{app:baselines}.

\subsubsection{Metrics}\label{sec:metrics} For offline experiments, we follow~\cite{fu2023wisdom,fu2024art} and adopt HR$_{seren}$@K, NDCG$_{seren}$@K, and MAP$_{seren}$@K, where K=1,3,5,10. HR$_{seren}$@K denotes the proportion of serendipity items retrieved in the top-K position. NDCG$_{seren}$@K is derived from the ranking metric NDCG, which replaces relevant items with serendipitous items. Similarly, MAP$_{seren}$@K is also a serendipity-version of MAP. 

In online experiments, there are three types of metrics: serendipity, utility, and user engagement metrics. \textbf{Serendipity} metrics focus on serendipity-related items. Here, it is slightly different from Definition~\ref{def:serend}, requiring only that items' atomic category, brand, or seller should not have appeared in the user's valid exposures during the last 10 days with visits. Some key metrics include:  \textbf{PVR} (the percentage of finally exposed items that belong to serendipity-related items), the number of \textbf{Clicks}, and \textbf{Transaction Volume} (number of purchases) for serendipity-related items. For simplicity, we refer to them as \textbf{S-PVR}, \textbf{S-Click}, and \textbf{S-TV}. Generally, the higher these metrics are, the better the model performs regarding serendipity.

The \textbf{utility} metrics measure the effectiveness and revenue of RSs across all items, including serendipitous and non-serendipitous ones. Common metrics include \textbf{CTR} (Click-Through Rate) and total \textbf{TV} (Transaction Volume) across all items. These metrics indicate the impact of increasing serendipity on the platform's performance and revenue.  The primary indicator of \textbf{user engagement} here is \textbf{UV3}, which measures the number of users who scroll through more than 200 items. An increase in UV3 suggests that users find the recommendations engaging and remain interested for longer, indicating an improvement in user experience.

% \subsubsection{Metrics} For \textbf{item prediction}, we follow~\cite{fu2023wisdom,fu2024art} and adopt HR$_{seren}$@K, NDCG$_{seren}$@K, and MAP$_{seren}$@K, where K=1,5,10. HR$_{seren}$@K denotes the proportion of serendipity items retrieved in the top-K position. NDCG$_{seren}$@K is derived from the well-known ranking metric NDCG, which replaces relevant items with serendipitous items. Similarly, MAP$_{seren}$@K is also a serendipity-version of MAP. For \textbf{search query prediction} task, we use \textit{hit rate}, the proportion of test samples where the generated query matches the ground truth query. As different queries may have the same meaning, we train an LLM-based relevance model to determine whether two queries are semantically aligned (matched).

\subsubsection{Reproducibility} During the training of SerenGPT, we employ \textbf{Qwen2-0.5B-Instruct}~\footnote{\url{https://huggingface.co/Qwen/Qwen2-0.5B-Instruct}} as the backbone LLM and split the training data into SFT and IPO training data at 1:1. In the SFT phase, SerenGPT is fully fine-tuned on 8 A100 GPUs. In the IPO phase, SerenGPT undergoes LoRA~\cite{hu2021lora} fine-tuning on 8 L40S GPUs, with a LoRA rank of 8, LoRA alpha set to 16, and LoRA dropout set to 0. Both phases use AdamW as the optimizer, with a batch size of 16, a learning rate of 1e-5, and the epoch set to 1. Additionally, the weight $\alpha$ of SFT loss in Eq.~\eqref{eq:final_ipo} is 0.2. During the inference, we utilize vLLM for acceleration, with a temperature of 0.95 and a repetition penalty of 1.05. The LLM-based baselines utilize the same backbone LLM as SerenGPT and are trained on the same training data.

In \textbf{online deployment}, SerenGPT is trained offline and then deployed online following Section~\ref{sec:downstream} with the nearline cache updated every 24 hours. Once deployed, SerenGPT is fixed. Only retrieval and pre-ranking models in the serendipity channel are continuously trained and updated at low costs. In practice, Taobao hashes daily visitors into 100 buckets, each serving millions of users. Cognition profile generation relies on the fine-tuned Qwen-7B-Chat, with profile updates every 15 days. For all users in a bucket, generating profiles on H20 takes about 480 GPU hours, while generating recommendations with SerenGPT on L40S takes 2400 GPU hours. 
% Generating profiles for all users in a single bucket takes about 6 hours on 80 H20 GPUs. Generating recommendations for all users in a bucket using SerenGPT takes about 50 hours on 48 L40S GPUs.

% \subsubsection{Reproducibility} During the training of both item and search query prediction tasks, we employ \textbf{Qwen2-0.5B-Instruct}~\footnote{\url{https://huggingface.co/Qwen/Qwen2-0.5B-Instruct}} as the backbone LLM for SerenGPT and split the training data into SFT and IPO training data at 1:1. In the SFT phase, SerenGPT is fully fine-tuned on 8 A100 GPUs. In the IPO phase, SerenGPT undergoes LoRA fine-tuning on 8 L40S GPUs, with a LoRA rank of 8, LoRA alpha set to 16, and LoRA dropout set to 0. Both phases use AdamW as the optimizer, with a batch size of 16, a learning rate of 1e-5, and the epoch set to 1. Additionally, we incorporate the SFT loss function in the IPO training with a weight of 0.1. During the inference, we utilize vLLM for acceleration, with a temperature of 0.95 and a repetition penalty of 1.05.

\vspace{-5pt}
\subsection{Offline Experiments}
As SERAL is an online framework, the offline experiments focus on SERAL's core component, SerenGPT, while the online experiments in Section~\ref{sec:online} presents the overall performance of SERAL.
\subsubsection{Offline Performance}
The performance is presented in Table~\ref{tab:overall}. Several important observations can be inferred from the results. 

\textit{Firstly, our model, SerenGPT, significantly outperforms baselines, especially in ranking and head metrics.} For example, SerenGPT surpasses the strongest baseline, SerenPrompt, by 226\% in Hit@1 and 79.5\% in NDCG@3. However, in tail recall metrics like HR@10, SerenGPT performs worse than SerenPrompt. This is likely because SerenPrompt scores every item in the candidate set, whereas SerenGPT only generates a single predicted item and then uses a matching model to match items in the candidate set. When the generated item is in the candidate set, it is matched accurately, boosting head metrics like Hit@1. However, when the generated item is not in the candidate set, the performance heavily relies on the matching model, which may struggle to recall relevant items, leading to poorer tail metrics. We opt for the generative model over the scoring model due to the high inference latency of LLMs, making online scoring of multiple items inefficient. Their inference efficiency is compared in the Appendix~\ref{app:efficiency}.

\textit{Secondly, LLM-based serendipity models perform consistently well and exhibit strong data efficiency.} Even zero-shot LLM4Seren outperforms most deep learning-based models, and fine-tuned LLMs further enhance performance owing to LLMs' stronger understanding of serendipity. Given the scarcity of serendipity data, deep learning-based models often rely on non-serendipity data for augmentation. In contrast, LLM-based serendipity models, SerenPrompt and SerenGPT, achieve superior results with only a small set of serendipity data. This highlights LLM-based models' effectiveness and data efficiency, making them suited for serendipity recommendations.

% \subsubsection{Search Query Prediction} Since there was no existing serendipity model for the task of search query prediction, we compare several baselines that we designed, which are based on the SerenGPT structure with modifications to the preference alignment approaches. These can also be seen as ablations of SerenGPT. The final results are presented in Table~\ref{tab:overall_search}, from which we can see that the SerenGPT trained with IPO performs the best. This highlights the importance of preference alignment and choosing the right alignment method. Some alignment methods, such as DPO, perform worse than SFT, which may be due to overfitting, as discussed in Appendix~\ref{app:diversity}, leading to the generation of overly homogeneous items.

% \begin{table}[h]

%     \vspace{-8pt}
%     \caption{Performance on search query prediction task.}
%     % \caption{Overall performance on search query prediction task.  The best result is given in bold, while the second-best value is underlined. The symbol * indicates statistically significant improvement over the best baselines( t-test with $p < 0.05$).}
%     \vspace{-8pt}
%     \centering
%     \scalebox{0.95}{
%     \setlength{\tabcolsep}{1.5mm}
%     {\begin{tabular}{c|cccccc}
% \toprule
% Method & ZSLLM & SFT & DPO & SimPO & KTO & \textbf{IPO(ours)} \\
% \midrule
% hit rate &  & 0.3298 & 0.3191 & 0.3289 &  & \textbf{0.3485*} \\
%  \bottomrule
% \end{tabular}}
% }
% \label{tab:overall_search}
% \vspace{-8pt}
% \end{table}
\subsubsection{Ablation Study}
To investigate the effectiveness of each module, we design several variants of SerenGPT. First, we remove some key modules: \textbf{w/o CP} excludes cognitive profiling from the prompt. \textbf{w/o CDI-D} and \textbf{w/o CDI-P} remove the denoising and pairing components of CDI, respectively, and \textbf{w/o CDI} removes both. Note that generating enough preference pairs is challenging without CDI pair, so w/o CDI pair employs KTO training~\cite{ethayarajh2024kto} that doesn't require pairs. \textbf{w/o IPO} discards IPO training, retaining only SFT on a subset of the data. \textbf{w/ $\alpha=0$} sets the weight $\alpha$ of SFT loss in Eq.~\eqref{eq:final_ipo} to 0. Additionally, we replace IPO with SFT and DPO, resulting in \textbf{w/ SFT} and \textbf{w/ DPO} variants.
\begin{table}[h]

    \vspace{-5pt}
    \caption{Ablation study of SerenGPT. }
    \vspace{-8pt}
    \centering
    \scalebox{0.9}{
    \setlength{\tabcolsep}{1.5mm}
    {\begin{tabular}{ccc|cc|cc}
\toprule
 \multirow{2}{*}{Method} & \multicolumn{2}{c|}{MAP$_{seren}$} & \multicolumn{2}{c|}{NDCG$_{seren}$} & \multicolumn{2}{c}{HR$_{seren}$} \\
 \cmidrule{2-7}
 & @3 & @10 & @3 & @10 & @3 & @10 \\
 \midrule
w/o CP & 0.2933 & 0.3195 & 0.3031 & 0.3558 & 0.4148 & 0.6450 \\
w/o CDI & 0.2840 & 0.3066 & 0.2924 & 0.3377 & 0.4103 & 0.6101 \\
w/o CDI-D & 0.2998 & 0.3224 & 0.3086 & 0.3542 & 0.4307 & 0.6313 \\
w/o CDI-P & \underline{0.3223} & \underline{0.3397} & \underline{0.3303} & 0.3653 & 0.4534 & 0.6101 \\
w/o IPO & 0.2198 & 0.2476 & 0.2312 & 0.2866 & 0.3316 & 0.5632 \\
\midrule
w/ $\alpha=0$ & 0.2887 & 0.3134 & 0.2969 & 0.3460 & 0.4080 & 0.6291 \\
w/ SFT & 0.3117 & 0.3335 & 0.3216 & \underline{0.3663} & \textbf{0.4640} & \underline{0.6616} \\
w/ DPO & 0.2926 & 0.3144 & 0.3027 & 0.3459 & 0.4239 & 0.6192 \\
\midrule
\textbf{SerenGPT} & \textbf{0.3260} & \textbf{0.3509} & \textbf{0.3353} & \textbf{0.3847} & \underline{0.4625} & \textbf{0.6760}\\ 
\bottomrule
\end{tabular}}
    }

\label{tab:Ablation}
\vspace{-5pt}
\end{table}

% \begin{table}[h]

%     \vspace{-5pt}
%     \caption{Ablation study of SerenGPT. }
%     \vspace{-8pt}
%     \centering
%     \scalebox{0.88}{
%     \setlength{\tabcolsep}{1.5mm}
%     {\begin{tabular}{ccc|cc|cc}
% \toprule
%  \multirow{2}{*}{Method} & \multicolumn{2}{c|}{MAP$_{seren}$} & \multicolumn{2}{c|}{NDCG$_{seren}$} & \multicolumn{2}{c}{NDCG$_{seren}$} \\
%  \cmidrule{2-7}
%  & @1 & @5 & @1 & @5 & @1 & @5 \\
%  \midrule
% w/o CP & 0.2513 & 0.3065 & 0.2513 & 0.3262 & 0.2513 & 0.5042 \\
% w/o CDI & 0.2445 & 0.2960 & 0.2445 & 0.3141 & 0.2445 & 0.4996 \\
% w/o CDI denoise & 0.2619 & 0.3110 & 0.2619 & 0.3284 & 0.2619 & 0.5193 \\
% w/o CDI pair & \underline{0.2860} & \underline{0.3317} & \underline{0.2860} & \underline{0.3470} & \underline{0.2860} & 0.5299 \\
% w/o IPO & 0.1771 & 0.2327 & 0.1771 & 0.2530 & 0.1771 & 0.4148 \\
% \midrule
% w/ SFT & 0.2695 & 0.3224 & 0.2695 & 0.3409 & 0.2695 & \underline{0.5496} \\
% w/ DPO & 0.2513 & 0.3031 & 0.2513 & 0.3210 & 0.2513 & 0.5042 \\
% \midrule
% SerenGPT & \textbf{0.2861} & \textbf{0.3395} & \textbf{0.2861} & \textbf{0.3587} & \textbf{0.2861} & \textbf{0.5556}\\ 
% \bottomrule
% \end{tabular}}
%     }

% \label{tab:Ablation}
% \vspace{-5pt}
% \end{table}

Due to page limits, we present representative results in Table~\ref{tab:Ablation}, with a full table in Appendix~\ref{app:ablation}. Removing any component leads to performance degradation, demonstrating the importance of each module. Notably, removing IPO causes the largest drop, highlighting the critical role of preference alignment. The significant decline in w/o CDI further confirms that our data intervention measures effectively enhance data quality. Lastly, replacing IPO with DPO or SFT reduces performance and generates more homogeneous content, whereas IPO increases diversity, as validated in Appendix~\ref{app:diversity}.

\subsection{Online Experiment}\label{sec:online}
\subsubsection{Short-term Online A/B Test}\label{sec:short_ab}
To evaluate the online impact of SERAL, particularly SerenGPT, we conduct a \textbf{two-week A/B test} on "Guess What You Like" of Taobao. We compare two variants, \textbf{SERAL-SFT} and \textbf{SERAL-IPO}, utilizing SerenGPT trained with SFT and IPO, respectively, against the online serendipity-enhanced baseline in Table~\ref{tab:short_ab}. Notably, the online baseline already incorporates cognitive profiles and trending topics to enhance serendipity, as these features were developed and deployed earlier. Since S-PVR and CTR are percentages, we use \textbf{pt} for absolute difference. For example, SerenGPT-IPO's +5.7pt in S-PVR represents an increase from 18.33\% to 24.03\%, an absolute improvement of 5.7\%.
% \begin{table}[h]
%     \vspace{-8pt}
%     \caption{Online improvement on serendipity metrics.}
%     \vspace{-8pt}
%     \centering
%     \scalebox{1}{
%     \setlength{\tabcolsep}{2mm}
%     {\begin{tabular}{c|ccc}
% \toprule
% Model & PVR & Clicks & Transaction Volume \\
% \midrule
% SerenGPT-SFT & +6.4\% & +33.97\% & +29.89\% \\
% SerenGPT-IPO & +6.4\% & +34.41\% & +30.66\% \\
%  \bottomrule
% \end{tabular}}
% }
% \label{tab:short_ab_seren}
% \vspace{-8pt}
% \end{table}
% 
\begin{table}[h]
    \vspace{-4pt}
    \caption{Online improvement over serendipity baseline. Since S-PVR and CTR are percentages, we use percentage points (pt) for absolute improvement. \% denotes relative improvement.}
    \vspace{-5pt}
    \centering
    \scalebox{0.87}{
    \setlength{\tabcolsep}{1.5mm}
    {\begin{tabular}{c|cccccc}
\toprule
% Model & S-PVR & S-Click & S-TV & CTR & TV & UV3 \\
% \midrule
% SerenGPT-SFT & +6.4\% & +33.97\% & +29.89\% & -0.41\% & -0.69\% & +0.54\% \\
% SerenGPT-IPO & +6.4\% & +34.41\% & +30.66\% & -0.20\% & -0.79\% & +0.25\% \\
Model & S-PVR & S-Click & S-TV & CTR & TV & UV3 \\
\midrule
SERAL-SFT & +5.68pt & +26.12\% & +24.58\% & -0.02pt & -0.65\% & +0.48\% \\
SERAL-IPO & +5.70pt & +29.56\% & +27.60\% & -0.01pt & -0.71\% & +0.14\% \\
 \bottomrule
\end{tabular}}
}
\label{tab:short_ab}
\vspace{-5pt}
\end{table}

Both variants of SERAL show significant improvements in serendipity metrics, such as exposure ratio (S-PRV), clicks (S-Click), and transaction volume (S-TV) of serendipity-related items. This shows that SERAL increased the proportion of serendipitous items in recommendations, and these items successfully captured user interest (clicks and purchases). It also improves the user engagement metric UV3, indicating that users are more engaged. Its impact on overall utility metrics, such as CTR and total transaction volume (TV), is relatively minor. Besides, SERAL-IPO outperforms SERAL-SFT in serendipity metrics, while maintaining comparable utility performance. This demonstrates that IPO can deliver serendipitous and engaging recommendations.
% \begin{table}[h]
%     \vspace{-8pt}
%     \caption{Improvement on utility and user experience metrics.}
%     \vspace{-8pt}
%     \centering
%     \scalebox{1}{
%     \setlength{\tabcolsep}{2mm}
%     {\begin{tabular}{c|ccc}
% \toprule
% Model & CTR & Transaction Volume & UV3 \\
% \midrule
% SerenGPT-SFT & -0.02\% & -0.69\% & +0.54\% \\
% SerenGPT-IPO & -0.01\% & -0.79\% & +0.25\% \\
%  \bottomrule
% \end{tabular}}
% }
% \label{tab:short_ab_util}
% \vspace{-8pt}
% \end{table}
\begin{figure}
    \centering
    \vspace{-10pt}
    \includegraphics[trim={0.3cm 0
    0 0},clip,width=0.49\textwidth]{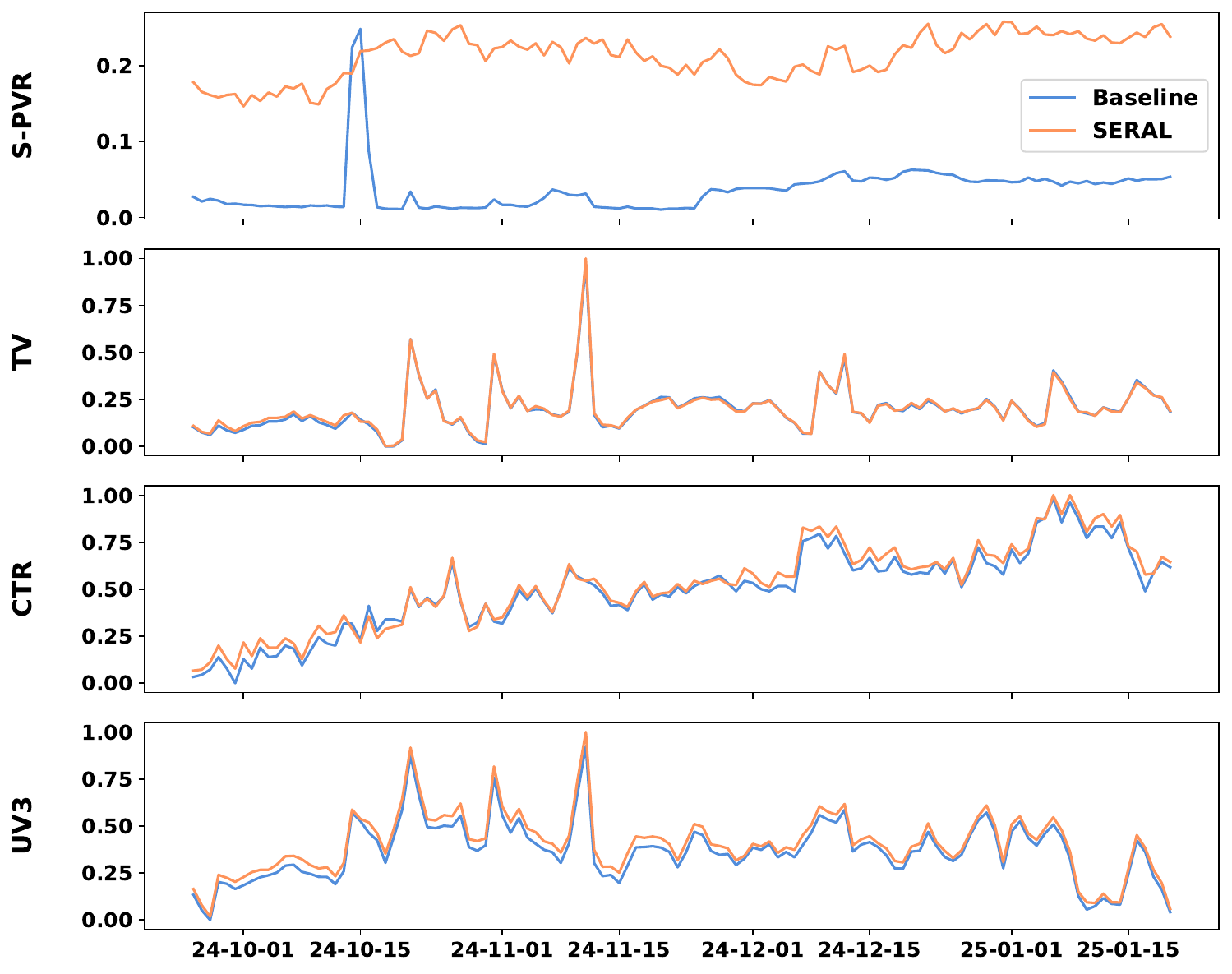}
    \vspace{-20pt}
    \caption{Long-term online impact of serendipity items.}
    \vspace{-10pt}
    \label{fig:long_ab}
\end{figure}
% \begin{figure}
%     \centering
%     % \vspace{-10pt}
%     \includegraphics[trim={0.3cm 0
%     0 0},clip,width=0.48\textwidth]{imgs/online_metrics.pdf}
%     \vspace{-20pt}
%     \caption{Long-term online impact of our SERAL.}
%     \vspace{-10pt}
%     \label{fig:long_ab}
% \end{figure}
\subsubsection{Long-term Online Impact of Serendipity Items}
Our proposed SERAL now serves the primary traffic in the "Guess What You Like" on Taobao. To investigate the long-term effects of serendipity items in large-scale RSs, we maintain a small traffic group without any serendipity enhancements as our baselines and compare it with SERAL for \textbf{four months}. Their daily performance and SERAL's average improvement are presented in Figure~\ref{fig:long_ab} and Table~\ref{tab:long_ab}. Note that, in Figure~\ref{fig:long_ab}, we apply min-max normalization to the y-axis of TV, CTR, and UV3 to protect sensitive data. For example, if the original data range is 1,000-5,000, it is scaled to 0-1. This only modifies the y-axis without altering trends and relative changes.
% More metrics are provided in Appendix~\ref{app:long_ab}.

\begin{table}[h]
    \vspace{-10pt}
    \caption{Online improvement over non-serendipity baseline. }
    \vspace{-8pt}
    \centering
    \scalebox{0.87}{
    \setlength{\tabcolsep}{1.5mm}
    {\begin{tabular}{c|cccccc}
\toprule
Model & S-PVR & S-Click & S-TV & CTR & TV & UV3 \\
\midrule
SERAL & +17.36pt & +819.9\% & +969.6\% & +0.05pt & +0.98\% & +3.04\% \\
 \bottomrule
\end{tabular}}
}
\label{tab:long_ab}
\vspace{-10pt}
\end{table}
% Firstly, serendipity model significantly increases the proportion of serendipitous items in recommendations. Over four months, the average \textbf{serendipity exposure ratio (S-PVR) rises from 3.35\% to 20.71\%} and serendipity-related clicks (S-Click) and transaction volumes (S-TV) also achieves substantial increase. Secondly, utility metrics (TV and CTR) for all items remain stable, with a slight increase of 0.98\% and 0.05pt. Lastly, there is a significant improvement in user engagement, with a \textbf{3.04\%} average increase in \textbf{UV3}. 
Compared to the baseline, our method has achieved a significant enhancement in serendipity items, reflected in S-PVR, S-Click, and S-TV. The increase of serendipitous items has bolstered the overall transaction volume (TV) and CTR by \textbf{0.98\%} and \textbf{0.05pt}, respectively. Moreover, it has also markedly improved user engagement, with a \textbf{3.04\%} average increase in \textbf{UV3}. This demonstrates that serendipity items help to break the filter bubble of homogeneous recommendations and effectively improve user engagement while also boosting CTR and revenue. The improvements here differ from those in Section~\ref{sec:short_ab} because various SERAL-related approaches are employed online, \eg, retrieval by SerenGPT and cognitive profiles. In contrast, Section~\ref{sec:short_ab} focuses solely on the impact of SerenGPT.

\subsubsection{Nearline Cache} 
\begin{figure}[h]
    \centering
    \vspace{-5pt}
    \includegraphics[trim={0.3cm 0cm
    0 2cm},clip,width=0.48\textwidth]{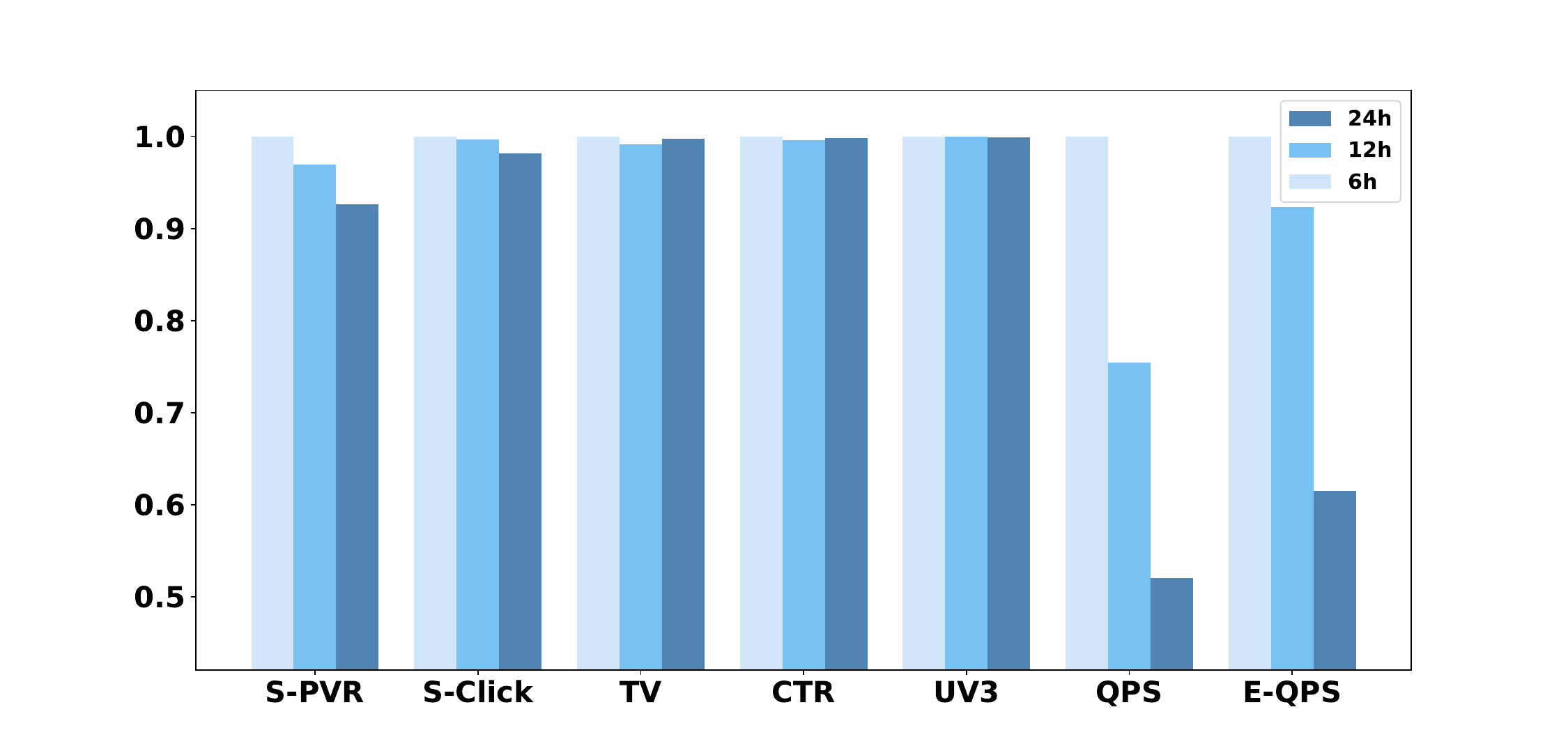}
    \vspace{-25pt}
    \caption{Online Performance of different caching frequency.}
    \vspace{-15pt}
    \label{fig:cache}
\end{figure}

 This section explores the impact of nearline caching frequency in a \textbf{two-week online test}, with the frequency set to \textbf{6h}, \textbf{12h}, and \textbf{24h}. The performance on different types of metrics (S-PVR, S-Click, CTR, TV, and UV3) and resource usage (\textbf{QPS} and \textbf{E-QPS}) are shown in Figure~\ref{fig:cache}. Here, QPS refers to queries per second for invoking SerenGPT, with higher QPS indicating greater resource usage. E-QPS denotes failed requests, and a higher E-QPS indicates greater resource strain, making it more difficult to access resources and provide recommendations. To facilitate plotting and protect sensitive data, all metrics are normalized to 1 based on their maximum values. As shown in Figure~\ref{fig:cache}, a higher cache frequency leads to better serendipity (S-PVR, S-Click), but it also results in greater resource consumption (QPS). Higher frequencies may lead to resource congestion, blocking the recommendation process and significantly reducing effectiveness. The cache frequency has little impact on the user experience metric UV3 and only a small effect on utility. However, even with a 24-hour update frequency, the improvement in serendipity remains significant. Therefore, the caching frequency should be adjusted based on available resources to strike a balance between performance and efficiency.

% \begin{figure}
%     \centering
%     \vspace{-10pt}
%     \includegraphics[trim={0.3cm 0
%     0 0},clip,width=0.48\textwidth]{imgs/freq_exp.pdf}
%     \vspace{-20pt}
%     \caption{Online Performance of different caching frequency .}
%     \vspace{-10pt}
%     \label{fig:cache}
% \end{figure}

% \textit{add the figure and other observations when experiments are over}

% \subsubsection{Compatibility Analysis}

% \subsubsection{Case Study} profile+user history behaviors+ serendipity item

\section{Conclusion}
This work addresses the filter bubble problem in RSs by proposing  SERAL. It comprises three stages: Cognition Profile Generation to compress user behavior into multi-level profiles, SerenGPT Alignment to align LLM-based serendipity predictions with human preferences, and Nearline Adaptation for efficient industrial deployment. Online experiments show it improves serendipity PVR, clicks, and transactions by 5.7\%, 29.56\%, and 27.6\%, enhancing user experience without much impact on revenue. Fully deployed in Taobao’s "Guess What You Like" section, it shows LLMs' potential to break the filter bubble and elevate user satisfaction in RSs.

\bibliographystyle{ACM-Reference-Format}
\bibliography{sample-base}

\balance
\newpage
\appendix
\section{Baselines}\label{app:baselines}
 For serendipity recommendations, we follow~\cite{fu2024art} and adopt three categories of baselines: sequential recommendation algorithms, deep learning-based serendipity recommendation algorithms, and LLM-based serendipity recommendation algorithms.

\textbf{Sequential recommendation} are trained on both serendipity and non-serendipity data. We selected two classic algorithms: 
\begin{itemize}
    \item \textbf{SASRec}~\cite{kang2018self} captures long-term semantics with a self-attention mechanism for sequential recommendation.
    \item \textbf{BERT4Rec}~\cite{sun2019bert4rec} introduces a Bidirectional Encoder Representation from Transformers for the sequential recommendation.
\end{itemize}
\textbf{Deep learning-based serendipity recommendation algorithms} are mainly trained on serendipity data, but due to the limited amount of such data, they leverage additional non-serendipity data for augmentation. In this category, we choose two state-of-the-art baselines: 
\begin{itemize}
    \item \textbf{SerenEhance}~\cite{fu2023wisdom} devises a self-enhanced module to learn fine-grained facets of serendipity for mitigating the data sparsity problem.
    % \item DESR~\cite{li2020directional} empolys serendipity vector to combine long-term preferences with short-term demands and generate serendipitous recommendations.
    \item \textbf{PURS}~\cite{li2020purs}  incorporates unexpectedness into recommendation by providing multi-cluster modeling of user interests and personalized unexpectedness via self-attention.
    % \item SNPR~\cite{zhang2021snpr} formulates serendipitous POI recommendation as a supervised multi-task learning problem and sloves it with Transformer.
\end{itemize}
\textbf{LLM-based serendipity recommendation algorithms} are trained only on serendipity data. This is a relatively new area, and we identify two baselines:
\begin{itemize}
    \item \textbf{LLM4Seren}~\cite{tokutake2024can} uses zero-shot LLMs to assess the serendipity of candidate items.
    \item \textbf{SerenPrompt}~\cite{fu2024art} employ various prompts and prompt tuning for LLMs to assess serendipity.
\end{itemize}
Our algorithm SerenGPT is also trained solely on serendipity data. While there are slight differences in training data, we ensure the use of the same test data for fair comparisons. Additionally, the LLM-based baselines utilize the same backbone LLM as SerenGPT and are trained on the same amount of training data.

\section{Diversity Analysis}\label{app:diversity} 
\begin{table}[ht]
\centering
% \vspace{-7pt}
\caption{Diversity comparison between different tuning strategies on serendipity recommendation. }
% \vspace{-5pt}
\scalebox{1}{
\setlength{\tabcolsep}{1.5mm}{
\begin{tabular}{cccc}
\toprule
Method & avg\_title\_cnt & avg\_cate\_cnt & hit\_rate \\
\midrule
SFT & 29.2 & 83.0 & 0.5483 \\
KTO & 23.7 & 25.3 & 0.4315 \\
DPO & 25.5 & 56.7 & 0.4845 \\
\textbf{IPO (ours)} & \textbf{30.0} & \textbf{106.8} & \textbf{0.5532} \\
\bottomrule
\end{tabular}
}}
% \vspace{-5pt}
\label{tab:diversity}
\end{table}

In our online experiments, we find that the diversity of recommendation produced by different preference alignment algorithms varies widely, and the IPO~\cite{azar2024general} algorithm generates more diverse results while maintaining accuracy. Therefore, we ultimately choose IPO as the preference alignment algorithm.

During the nearline adaptation step, the predictions generated by SerenGPT are used to retrieve items from the candidate item pool. Specifically, we perform 30 sampling iterations for each user, generating 30 titles, which are then used to retrieve items across different categories. We evaluate the results generated by SFT, DPO~\cite{rafailov2024direct}, KTO~\cite{ethayarajh2024kto}, and IPO~\cite{azar2024general} on the test set by analyzing three metrics: \textit{avg\_title\_cnt} (number of unique titles among 30 samples), \textit{avg\_cate\_cnt} (the average number of categories retrieved), and \textit{hit\_rate} (whether the target serendipity item's category is included in the retrieved categories). 

From Table~\ref{tab:diversity}, we observe that IPO generates almost no duplicate titles, retrieves a wide variety of categories, and achieves a higher hit rate. This indicates that IPO can generate more diverse results while maintaining relevance, which is beneficial for producing serendipitous outcomes. We notice content homogeneity after the initial SFT stage, which hinders the generation of meaningful preference pairs. Initially, we bypass preference pair generation and directly adopt the pair-independent KTO approach, but this further amplifies the homogeneity (e.g., reduced number of titles and retrieved categories). To address this, we use CDI to generate higher-quality preference pairs for DPO, which alleviates the diversity issue but still underperforms SFT. Ultimately, IPO proves to be the best approach, achieving both higher accuracy and diversity. This improvement might be because SFT, KTO, and DPO tend to overfit when handling deterministic or near-deterministic data, whereas IPO effectively avoids such issues~\cite {azar2024general}.

\section{Efficiency Analysis}\label{app:efficiency}
To evaluate the inference efficiency of our model, we compare it with SOTA deep learning-based baseline SerenEnhance, two LLM-based models (LLM4Seren and SerenPrompt). We measure the average inference time each model takes to score a candidate set of 100 items, as shown in Table~\ref{tab:inference}. 

First, the deep learning-based model SerenEnhance has high inference efficiency, making it suitable for online serving. However, the LLM-based models have significantly longer inference times. To address this, we adopt a nearline caching approach to avoid high inference delays. Second, SerenGPT demonstrates substantially lower inference latency than the other LLM-based baselines. This is because SerenGPT is a generative model that directly predicts the next serendipitous item, bypassing the need to score each candidate individually. In contrast, LLM4Seren and SerenPrompt must score every candidate item, resulting in extremely long inference times. Additionally, their approach cannot support nearline caching since they require real-time scoring of candidate items, making it challenging to deploy them in production. This difference in efficiency highlights why we chose a generative approach with SerenGPT -- it reduces latency and enables practical deployment with nearline caching.

\begin{table}[ht]
\centering
\vspace{-5pt}
\caption{Comparison of average inference time. }
\vspace{-5pt}
\scalebox{0.9}{
\setlength{\tabcolsep}{1.5mm}{
\begin{tabular}{ccccc}
\toprule
Method & {SerenEnhance} & {LLM4Seren} & {SerenPrompt} & {SerenGPT} \\
\midrule
Time (s) & {0.0204} & {35.65} & {35.36} & {4.48} \\
\bottomrule
\end{tabular}
}}
\vspace{-5pt}
\label{tab:inference}
\end{table}

% \section{Case Study}
% To explore the specific recommendations made of our model and the online baseline, we visualize the recommendations for the same user during the same recommendation session in Figure~\ref{fig:case_study}. The user's historical behavior primarily involves winter women's clothing. In this context, the baseline model mostly recommends winter women's clothing, with only a few items outside this category. In contrast, our model leverages the user's profile and reasoning on historical behaviors to generate more insightful recommendations. For instance, the user profile indicates they own pets and enjoy nail art, so our model recommended pet care products and nail art supplies. Additionally, the user's recent interactions involve winter clothing, such as sweaters, which are prone to pilling. Therefore, our model also recommended a fabric shaver. This demonstrates that our model can utilize powerful reasoning capabilities to uncover user needs and provide more novel and useful recommendations.

% \begin{figure}
%     \centering
%     \vspace{-10pt}
%     \includegraphics[trim={0 5cm
%     0 0},clip,width=0.5\textwidth]{imgs/case_study-2.pdf}
%     \vspace{-20pt}
%     \caption{Case Study.}
%     \vspace{-10pt}
%     \label{fig:case_study}
% \end{figure}

% \section{Long-term Impact of Serendipity}\label{app:long_ab}

\section{Full Ablation Study}\label{app:ablation}
Due to space constraints, we present a subset of ablation metrics in the main text, with the full set of metrics available in Table~\ref{tab:Ablation_full}.

\begin{table*}[h]

    % \vspace{-10pt}
    \caption{Ablation study of SerenGPT. The best result is given in bold, while the second-best value is underlined. }
    % \vspace{-8pt}
    \centering
    \scalebox{1}{
    \setlength{\tabcolsep}{1.5mm}
    {\begin{tabular}{ccccc|cccc|cccc}
\toprule
\multirow{2}{*}{Method} & \multicolumn{4}{c|}{MAP$_{seren}$} & \multicolumn{4}{c|}{NDCG$_{seren}$} & \multicolumn{4}{c}{HR$_{seren}$} \\
\cmidrule{2-13}
 & @1 & @3 & @5 & @10 & @1 & @3 & @5 & @10 & @1 & @3 & @5 & @10 \\
 \midrule
w/o CP & 0.2513 & 0.2933 & 0.3065 & 0.3195 & 0.2513 & 0.3031 & 0.3262 & 0.3558 & 0.2513 & 0.4148 & 0.5042 & 0.6450 \\
w/o CDI & 0.2445 & 0.2840 & 0.2960 & 0.3066 & 0.2445 & 0.2924 & 0.3141 & 0.3377 & 0.2445 & 0.4103 & 0.4996 & 0.6101 \\
w/o CDI denoise & 0.2619 & 0.2998 & 0.3110 & 0.3224 & 0.2619 & 0.3086 & 0.3284 & 0.3542 & 0.2619 & 0.4307 & 0.5193 & 0.6313 \\
w/o CDI pair & \underline{0.2860} & \underline{0.3223} & \underline{0.3317} & \underline{0.3397} & \underline{0.2860} & \underline{0.3303} & \underline{0.3470} & 0.3653 & \underline{0.2860} & 0.4534 & 0.5299 & 0.6101 \\
w/o IPO & 0.1771 & 0.2198 & 0.2327 & 0.2476 & 0.1771 & 0.2312 & 0.2530 & 0.2866 & 0.1771 & 0.3316 & 0.4148 & 0.5632 \\
\midrule
w/ $\alpha=0$ & 0.2521 & 0.2887 & 0.3009 & 0.3134 & 0.2521 & 0.2969 & 0.3176 & 0.3460 & 0.2521 & 0.4080 & 0.4943 & 0.6291 \\
w/ SFT & 0.2695 & 0.3117 & 0.3224 & 0.3335 & 0.2695 & 0.3216 & 0.3409 & \underline{0.3663} & 0.2695 & \textbf{0.4640} & \underline{0.5496} & \underline{0.6616} \\
w/ DPO & 0.2513 & 0.2926 & 0.3031 & 0.3144 & 0.2513 & 0.3027 & 0.3210 & 0.3459 & 0.2513 & 0.4239 & 0.5042 & 0.6192 \\
\midrule
\textbf{SerenGPT} & \textbf{0.2861} & \textbf{0.3260} & \textbf{0.3395} & \textbf{0.3509} & \textbf{0.2861} & \textbf{0.3353} & \textbf{0.3587} & \textbf{0.3847} & \textbf{0.2861} & \underline{0.4625} & \textbf{0.5556} & \textbf{0.6760} \\ 

\bottomrule
\end{tabular}}
    }

\label{tab:Ablation_full}
% \vspace{-5pt}
\end{table*}

\section{Prompts and Examples}\label{app:prompt}
\begin{figure*}  
    \centering
    \includegraphics[width=\textwidth]{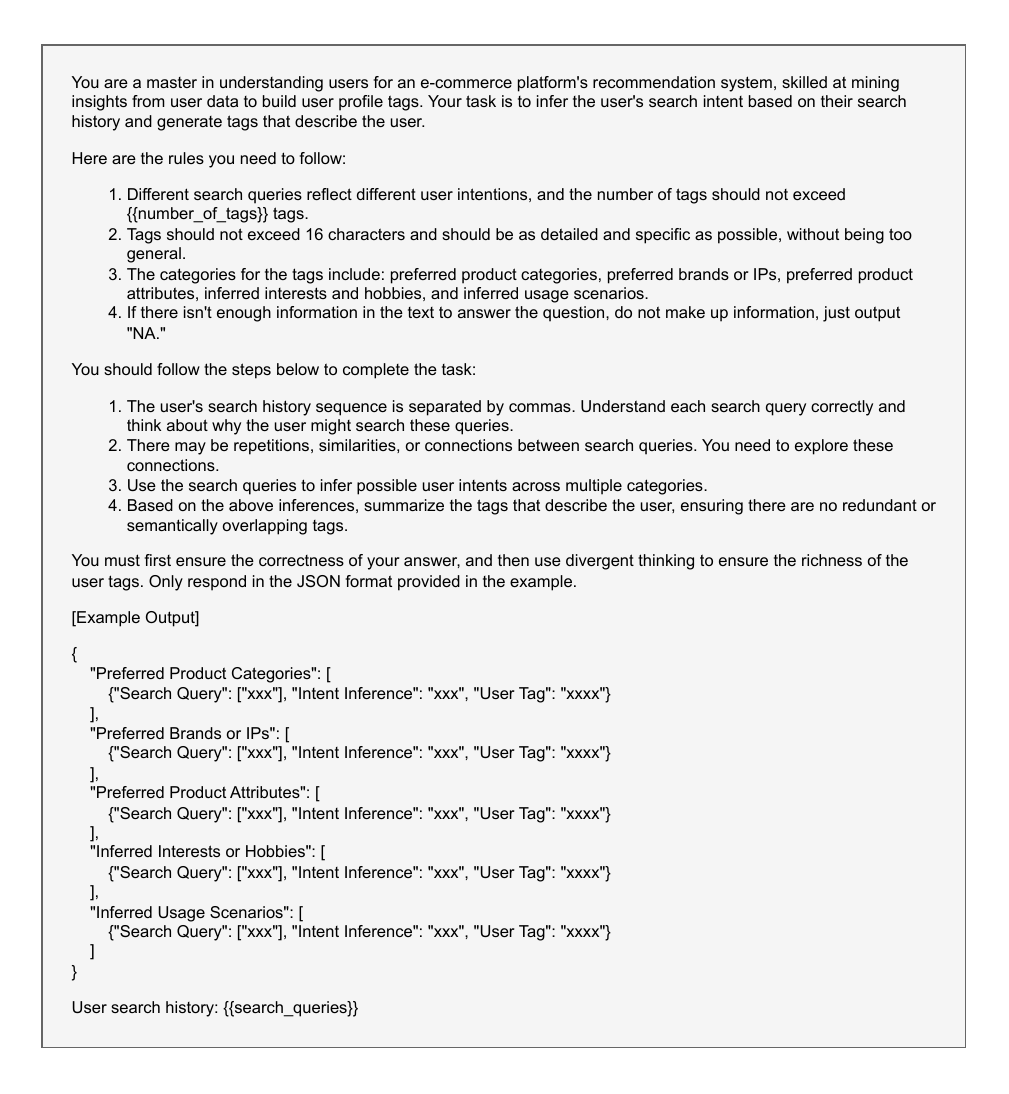}  
    % \vspace{-30pt}
    \caption{Prompt for short-term profile.}
    \label{fig:short-prompt}
\end{figure*}

The prompt for the short-term profile is long, so we provide only a brief description in the main text and place the complete prompt in Figure~\ref{fig:short-prompt}. The prompt for the long-term profile and SerenGPT is simpler, and we have described it in Section~\ref{sec:long-term} and~\ref{sec:sft}. 

% Here, we provide an example of SerenGPT's input (three types of profiles and user history) and its recommendations in Figure~\ref{fig:full-example}. To protect user privacy, we have withheld certain information in this example.
% The prompt for the short-term profile is long, so we provide only a brief description in the main text and place the complete prompt in Figure~\ref{fig:short-prompt}. The prompt for the long-term profile is simpler, and we have described it in Section~\ref{sec:long-term}. Here, we provide specific examples of three types of profiles: static, short-term, and long-term profiles, which can be found in Figure~\ref{fig:profile-example}. The prompt for SerenGPT is also described in Section~\ref{sec:sft} of the main text, and here we provide its recommendation example in Figure~\ref{fig:serenGPT-example}.

% \begin{figure} 
%     \centering
%     % \vspace{-20pt}
%     \includegraphics[trim={0.5cm 0.6cm
%     0 0.9cm},clip,width=0.5\textwidth]{imgs/full-example_v2.pdf}  
%     \vspace{-20pt}
%     \caption{Example of SerenGPT's input and recommendation.}
%     \label{fig:full-example}
% \end{figure}

% \begin{figure} 
%     \centering
%     \includegraphics[width=0.5\textwidth]{imgs/profile-example.pdf}  
%     \vspace{-30pt}
%     \caption{Example of cognition profile.}
%     \label{fig:profile-example}
% \end{figure}

% \begin{figure} 
%     \centering
%     \includegraphics[width=0.5\textwidth]{imgs/SerenGPT-example.pdf}  
%     \vspace{-30pt}
%     \caption{Example of SerenGPT's recommendation.}
%     \label{fig:serenGPT-example}
% \end{figure}

\section{Another Task: Search Query Prediction}
\subsection{Setup} In addition to utilizing SerenGPT for item title generation, we also explore another task, \textbf{search query prediction}, to enhance the serendipity recommendation in the "Guess What You Like" column of the Taobao App homepage. This task predicts the next serendipitous search query based on the user's historical behavior and profile. Users' search queries can also contain rich serendipity information, such as new preferences and unmet needs in current recommendations. Therefore, we filter out search queries that lead to serendipitous clicks and train SerenGPT to predict these queries. In the online serving, the predicted search queries will enter the recall phase of the serendipity channel just like the predicted items, ultimately generating recommendations. For offline experiments, we filter out 26, 672 serendipity queries and divide them into a training set and a test set in a 9:1 ratio, focusing on comparing the accuracy of predicted queries.

While the field of \textbf{search query prediction} has not specifically focused on serendipity, predicting search queries that users may find serendipitous is still highly valuable in search. It can help capture user interest, improve retention, and enhance satisfaction. Since this field heavily relies on textual information, traditional serendipity recommendation algorithms based on IDs are difficult to adapt. Therefore, we design several LLM-based baseline algorithms inspired by SerenGPT, particularly exploring various DPO variants, which also serve as an ablation study for our method:
\begin{itemize}
    \item \textbf{ZSLLM} employs LLM to predict the next serendipitous search query in a zero-shot setting.
    \item \textbf{SFT} refers to LLM being fine-tuned in a supervised manner on the full training dataset.
    \item \textbf{DPO} replaces SerenGPT's IPO~\cite{azar2024general} with DPO~\cite{rafailov2024direct}, a classic preference optimization algorithm based on pairwise preferences.
    \item \textbf{KTO} replaces SerenGPT's IPO with KTO~\cite{ethayarajh2024kto}, which relies solely on binary feedback indicating whether an output is desirable or undesirable.
    \item \textbf{SimPO} replaces SerenGPT's IPO with SimPO~\cite{meng2024simpo}, which eliminates the need for a reference model.
\end{itemize}
All these baselines share the same backbone LLM and use the same prompt as SerenGPT. An example prompt structure is: "\textit{Based on the user's profile \{\{cognitive profile\}\} and recent behavior history \{\{history\}\}, predict the next search query that the user might find serendipitous: }" 
% \muyan{user's profile \{cognitive profile\} and recent behavior history \{history\} ?}

For evaluation, we use \textit{hit rate}, the proportion of test samples where the generated query matches the ground truth query. Different queries may have the same meaning, so we train an LLM-based relevance model to determine whether two queries are semantically aligned (matched). 

\subsection{Offline and Online Performance}
Since there is no existing serendipity model for the task of search query prediction, we compare several baselines that we design, which are based on the SerenGPT structure with modifications to the preference alignment approaches. These can also be seen as ablations of SerenGPT. The final results are presented in Table~\ref{tab:overall_search}, from which we can see that the SerenGPT trained with IPO performs the best. This highlights the importance of preference alignment and choosing the right alignment method. Some alignment methods, such as DPO, perform worse than SFT, which may be due to overfitting, as discussed in Appendix~\ref{app:diversity}, leading to the generation of overly homogeneous items.

\begin{table}[h]

    % \vspace{-8pt}
    \caption{Performance on search query prediction task.}
    % \caption{Overall performance on search query prediction task.  The best result is given in bold, while the second-best value is underlined. The symbol * indicates statistically significant improvement over the best baselines( t-test with $p < 0.05$).}
    \vspace{-8pt}
    \centering
    \scalebox{0.95}{
    \setlength{\tabcolsep}{1.5mm}
    {\begin{tabular}{c|cccccc}
\toprule
Method & ZSLLM & SFT & DPO & SimPO & KTO & \textbf{IPO(ours)} \\
\midrule
hit rate & 0.2842  & 0.3298 & 0.3191 & 0.3289 &  0.2181 & \textbf{0.3485*} \\
 \bottomrule
\end{tabular}}
}
\label{tab:overall_search}
\vspace{-8pt}
\end{table}

We also conduct an A/B test on the search query prediction task, comparing our approach to the online serendipity baseline. We select search queries that led to serendipitous clicks and train SerenGPT offline to predict these queries. During online serving, the predicted search queries, similar to predicted items, are incorporated into the serendipity channel’s recall stage, ultimately contributing to serendipitous recommendations. In the A/B test, we observe that the exposure, clicks, and transaction volume of serendipity-related items increased by 6.34\%, 18.69\%, and 30.20\%, respectively. Additionally, the CTR improves by 0.19 percentage points. However, UV3 shows a slight decline of 1.08\%.

\end{document}